# A standardised protocol for assessment of relative SARS-CoV-2 variant severity, with application to severity risk for COVID-19 cases infected with Omicron BA.1 compared to Delta variants in six European countries

Tommy Nyberg[1], Peter Bager[2], Ingrid Bech Svalgaard[2], Dritan Bejko[3], Nick Bundle[4], Josie Evans[5], Tyra Grove Krause[2], Jim McMenamin[5], Joël Mossong[3], Heather Mutch[5], Ajibola Omokanye[4], André Peralta-Santos[6], Pedro Pinto Leite[6], Jostein Starrfelt[7], Simon Thelwall[8], Lamprini Veneti[7], Robert Whittaker[7], John Wood[5], Richard Pebody[9*], Anne M Presanis[1*]

**Affiliations**
1. MRC Biostatistics Unit, University of Cambridge, Cambridge, United Kingdom
2. Statens Serum Institut, Copenhagen, Denmark
3. Luxembourg Health Directorate, Luxembourg, Luxembourg
4. European Centre for Disease Prevention and Control, Stockholm, Sweden
5. Public Health Scotland, Glasgow, Scotland, United Kingdom
6. Directorate-General of Health, Lisbon, Portugal
7. Norwegian Institute of Public Health, Oslo, Norway
8. COVID-19 Vaccines and Epidemiology Division, UK Health Security Agency, London, United Kingdom
9. World Health Organization Regional Office for Europe, Copenhagen, Denmark
* These authors contributed equally to this work and share last authorship.

**Correspondence**: Tommy Nyberg, MRC Biostatistics Unit, University of Cambridge, East Forvie Building, Forvie Site, Robinson Way, Cambridge CB2 0SR, United Kingdom.
tommy.nyberg@mrc-bsu.cam.ac.uk

Abstract
Several SARS-CoV-2 variants that evolved during the COVID-19 pandemic have appeared to differ in severity, based on analyses of single-country datasets. With decreased SARS-CoV-2 testing and sequencing, international collaborative studies will become increasingly important for timely assessment of the severity of newly emerged variants. The Joint WHO Regional Office for Europe and ECDC Infection Severity Working Group was formed to produce and pilot a standardised study protocol to estimate relative variant case-severity in settings with individual-level SARS-CoV-2 testing and COVID-19 outcome data during periods when two variants were co-circulating. To assess feasibility, the study protocol and its associated statistical analysis code was applied by local investigators in Denmark, England, Luxembourg, Norway, Portugal and Scotland to assess the case-severity of Omicron BA.1 relative to Delta cases. After pooling estimates using meta-analysis methods (random effects estimates), the risk of hospital admission (adjusted hazard ratio [aHR]=0.41, 95% CI 0.31-0.54), ICU admission (aHR=0.12, 95% CI 0.05-0.27), and death (aHR=0.31, 95% CI 0.28-0.35) was lower for Omicron BA.1 compared to Delta cases. The aHRs varied by age group and vaccination status. In conclusion, this study has demonstrated the feasibility of conducting variant severity analyses in a multinational collaborative framework. The results add further evidence for the reduced severity of the Omicron BA.1 variant.

# Introduction

During the COVID-19 pandemic, several variants of the SARS-CoV-2 virus evolved that have differed in terms of the risk of causing severe disease [1-18]. A wide range of estimates of the relative risks associated with new variants have been reported, often with moderate precision. Some countries have published variant severity assessments based on national data, but many countries have not. In those countries that have not reported, there are likely observational data available on test-positive COVID-19 cases and their outcomes that could further inform about variant severity.

During the spring and summer of 2022, mass testing for COVID-19 as well as sequencing capacity was reduced in many countries where the pandemic appeared to recede [19-22]. As data on test-positive cases become less available, collaborative international efforts will become more important to rapidly identify differences in severity between virus variants.

To address these issues, the Joint WHO Regional Office for Europe and ECDC Infection Severity Working Group was formed with the aim of producing a standardised protocol that could be used by investigators in individual countries to analyse local available variant severity data using comparable methods and definitions. This decentralized approach overcomes potential issues in sharing individual-level data between countries. The intended application of the standardised protocol is the comparison of the risks between two virus variants, during calendar periods when both variants are circulating, and there are individual-level data on which virus variant caused the infection of each case. Outcomes include indicators of severe disease, such as hospital admission, intensive care unit (ICU) admission or death. The objective was to assess differences in SARS-CoV-2 variant severity, but the approach is applicable also to the study of severity of other similar diseases, such as influenza subtypes. This approach enables direct contrasting of relative risk estimates from several countries whose data have been analysed in a consistent way, and also allows international organisations and researchers to pool those estimates. Pooled estimates can provide greater precision due to a larger sample size compared with estimates from each separate country.

In this report, we describe the development of the standardised protocol. We then apply the protocol in a pilot study based on analysis of data on the Delta (B.1.617.2) and Omicron (B.1.1.529) BA.1 variants from six European countries.

# Methods

### Protocol development
The development of the study protocol and the associated statistical analysis code is summarised in Figure 1A. A first version of the protocol was drafted and circulated for comments from the countries participating in the working group, together with a survey where representatives from each country reported on the testing, variant classification and outcomes data available in each country. This survey and ensuing discussions resulted in a second revised protocol with further clarifications. After the second revised protocol had been circulated and approved, the country-level statistical analysis described in the protocol was translated into a standardised statistical analysis code, available at: https://github.com/TommyNyberg/variant_severity.

**Figure 1. Protocol development and application summary.** Schematic summary of (A) the development of the study protocol and standardised statistical analysis code, and (B) the application of the protocol and analysis code to estimate differences in COVID-19 severity outcomes for cases infected with the Omicron BA.1 compared to the Delta variant in six European countries.

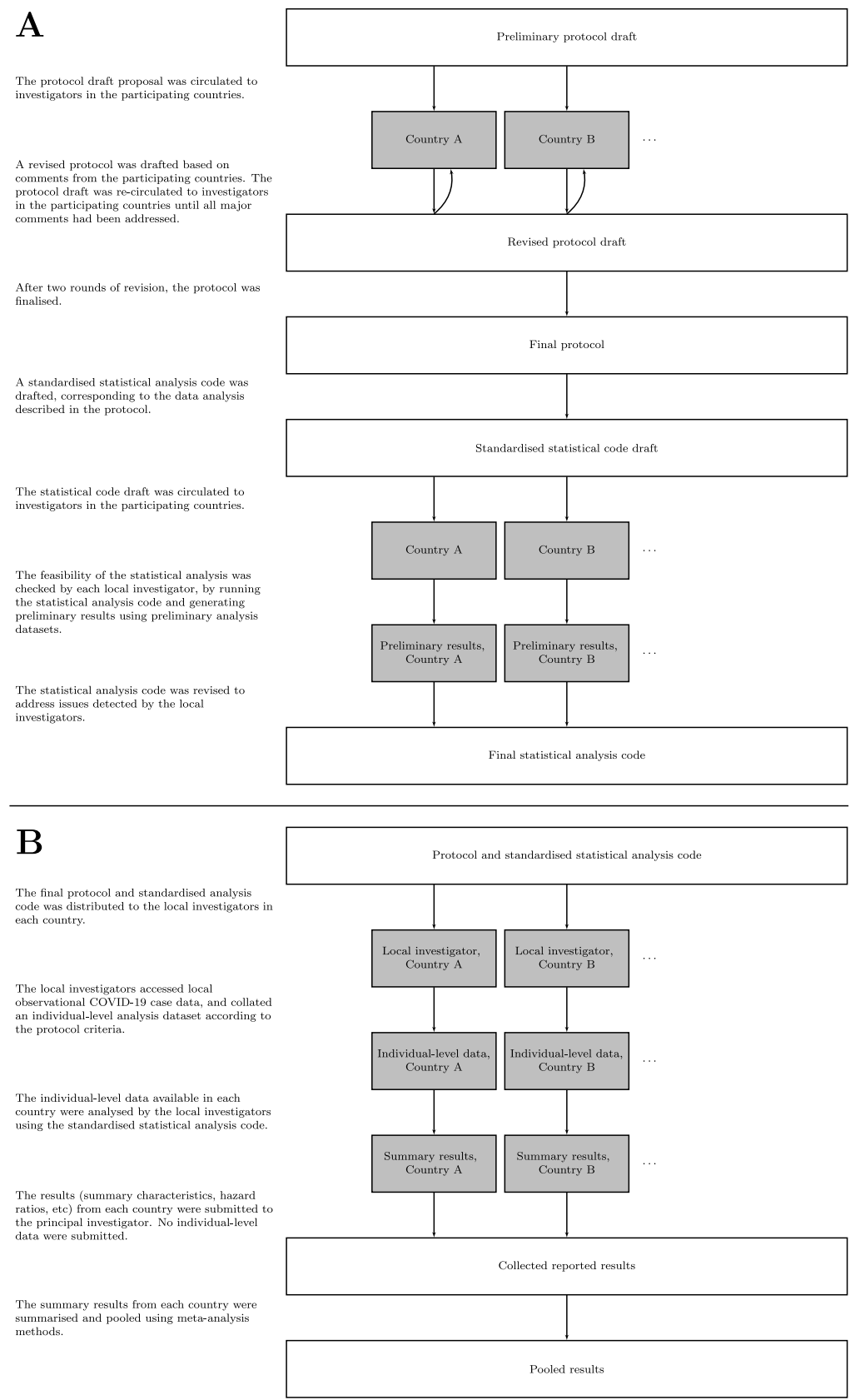

Data analysis

In a pilot study, the feasibility of performing the standardised analysis according to the study protocol was assessed. For this, each country used the statistical analysis code to analyse their national data on identified cases of the Delta or Omicron BA.1 variants.

The full protocol is available in the Supplementary material (Supplement A). In brief, retrospective cohorts of test-positive COVID-19 cases infected with Delta or Omicron BA.1 were identified by the local investigators. It was required that individual-level data to classify the SARS-CoV-2 variant of each case were available, for example through whole genome sequencing of swab samples taken for PCR testing, or through more rapid variant classification methods such as preliminary genotyping [23] or S gene status. It was recommended that the inclusion period be restricted to a period when there were cases available with both variants. If variant classification methods with less than 100% sensitivity and specificity such as S gene target failure were used, it was further recommended to restrict the inclusion period to the earliest and latest calendar date when variant prevalence was sufficiently high to ensure positive predictive values ≥90% for both variants. Otherwise, the choice of inclusion period was left to the discretion of the local investigators.

Hazard ratios (HRs) of severe outcomes following a positive test with Omicron BA.1 compared to Delta were estimated with Cox proportional hazards regression models, stratified by vaccination status, region and calendar time of positive test, and adjusted for age and sex at a minimum ("required adjustment variables"). For calendar time, two adjustments were considered, one where the stratification was for exact date, and one where the stratification was for calendar week with additional regression adjustment for exact date within each week. If available, adjustments for reinfection status ("highly desired adjustment variable"), and ethnicity, socio-economic status, comorbidity and international travel ("desired adjustment variables") were carried out in further analyses. Each of three severe outcomes, where available, were considered: hospital admission within 14 days of a positive test; ICU admission within 14 days of a positive test; and death within 28 days of a positive test. When available, the investigators were requested to provide results for both outcome events *due to* COVID-19 (COVID-19-specific outcomes) and among people *with* COVID-19 (all-cause outcomes), according to national criteria. Sub-group analyses by age group and vaccination status were also requested.

As a sensitivity analysis, the protocol further requested assessing the potential impact of epidemic phase bias, a bias that may occur when comparing two virus variants in different phases of growth or decline [24].

Meta-analysis

Using meta-analysis methods, the HR estimates from each country were contrasted and the heterogeneity between estimates was assessed using the $I^2$ statistic. If countries had reported HRs of COVID-19-specific outcomes these were used in the meta-analysis, and otherwise HRs of all-cause outcomes. The HRs were pooled using both fixed effects and random effects models.

The country-level analyses and the meta-analysis were performed using R software (R Foundation for Statistical Computing, Vienna, Austria), using the *survival*, *meta* and *forestploter* libraries.

## Implementation of the protocol

The protocol was applied as specified to all participant countries' available datasets, with one exception: in Denmark, during the Omicron BA.1 outbreak, reporting of Omicron variant classifications was prioritised over other variants for hospitalised cases, so that for some hospitalised cases with genotyping data it was only known whether they had Omicron or another variant. Since Delta was the predominant non-Omicron variant during the calendar period, cases of non-Omicron variants were assumed to be Delta cases. In a sensitivity analysis, the impact on the results of restricting to protocol-compliant Delta cases in Denmark whose Delta variant had been reported was explored.

The protocol specified events due to any cause as primary outcomes, and COVID-19-specific events as secondary outcomes. However, because not all participating countries were able to access data on all-cause events, the meta-analysis deviated from the protocol by instead using combined outcomes where HRs of COVID-19-specific events were used when available, and otherwise HRs of all-cause events.

The protocol specified additional subgroup analyses by two variables, to estimate vaccination-and-reinfection-status-specific and vaccination-status-and-age-group-specific adjusted HRs for Omicron BA.1 versus Delta. However, it was found that such subgroups were too small in countries with moderate sample sizes to reliably estimate these HRs. The results of these subgroup analyses were therefore not included.

# Results

## Protocol development

After circulation of the first draft of the protocol, a need was identified to simplify the original inclusion criteria which outlined circumstances under which variants classified through different methods were eligible: an initial requirement to choose the inclusion period based on a required minimum prevalence of each compared variant was removed in favour of a recommendation to choose a consecutive period when cases of each variant were prevalent, at the discretion of the local investigators. For the potential confounders, the revised protocol acknowledged that vaccination status should be defined to reflect the vaccination status categories (in terms of number of vaccine doses and time since last dose) applicable in the considered study inclusion period.

## Data analysis pilot

The results of the first round of the applied analysis resulted in further revisions. It was found that the analyses by age group or vaccination status had convergence issues when applied to data from some countries with small case numbers. The age groups and vaccination status groups were therefore redefined to ensure sufficient numbers of events in each category. Vaccination status was further revised to account for the potential effect of waning vaccine-induced immunity, by including a separate category for cases who had received two vaccine doses but where the second dose was received more than five months (≥153 days) before the positive test. The pilot analysis also identified that the results for countries with smaller datasets were sensitive to the choice of the number of regions to use in adjustments; it was recommended that up to approximately five regions were identified for countries with few cases, or not to use subnational regions for geographically small countries. It was further recognised that guidance should be provided on how to select the inclusion period: further instructions were circulated, ensuring that both Delta and Omicron cases were available each calendar week. Finally, minor code errors were identified and resolved, and code was added to perform additional checking of input data.

## Omicron BA.1 versus Delta variant severity

After the revision of the protocol and analysis code, the countries re-analysed their datasets using the final updated code (Figure 1B).

A detailed description of each country's data, including COVID-19 testing, variant classification procedures and data linkages, is available in the Supplementary material (Supplementary text). The supplementary material also shows inclusion summaries (Supplementary Figures S1-S6), a summary of available outcomes and adjustment variables (Supplementary Table S1), and descriptive frequencies of outcome events (Supplementary Table S2) and case characteristics (Supplementary Table S3), by country.

Five of the six countries analysed data on cases diagnosed at any age, with mean age ranging between 30.5 and 33.8 years by country. One country (Portugal) included only cases aged ≥16 years, with mean age 40.7 years. Two countries included only cases classified by whole genome sequencing or genotyping, while four countries additionally included cases whose variants were assessed by S gene target failure. The inclusion periods differed somewhat between countries according to differences in when Omicron superseded Delta as the dominant circulating variant, in line with the study protocol. By country, the proportion unvaccinated ranged from 11% to 39%, and the proportion that had received a booster vaccine dose ranged from 2.5% to 22%.

### Hazard ratios

Figures 2-4 show forest plots of adjusted HRs of hospital admission, ICU admission or death, based on the two adjustment strategies. Supplementary Figure S7 shows the corresponding unadjusted HRs. The HRs from all countries indicated lower risks of all severity outcomes for cases infected with Omicron BA.1 compared to Delta, both unadjusted and after adjustment for the required variables. Adjustment for the extended set of highly desired or desired adjustment variables changed the HRs only marginally compared to the HRs adjusted for the required adjustment variables only. The pooled random-effects HRs adjusted for the maximum number of adjustment variables were 0.41 (95% CI 0.31-0.54) for hospital admission, 0.12 (95% CI 0.05-0.27) for ICU admission, and 0.31 (95% CI 0.28-0.35) for death.

**Figure 2:** Hazard ratios of hospital admission (COVID-19-specific, where available, or otherwise due to any cause) for COVID-19 cases infected with the Omicron BA.1 versus Delta variants, by country and pooled across countries using a fixed effect and a random effect model.

(A). Adjusted for the required set of adjustment variables, using stratification for exact calendar date.
(B). Adjusted for the required and highly desired set of adjustment variables, using stratification for exact calendar date.
(C). Adjusted for the required, highly desired and desired set of adjustment variables, using stratification for exact calendar date.
(D). As (A), but using stratification for calendar week and regression adjustment for exact calendar date.
(E). As (B), but using stratification for calendar week and regression adjustment for exact calendar date.
(F). As (C), but using stratification for calendar week and regression adjustment for exact calendar date.

The required adjustment variables were: sex, age at diagnosis, calendar date of positive test, area of residence, vaccination status (all available for all countries). The highly desired adjustment variables were: reinfection status (available for all countries). The desired adjustment variables were: ethnicity/country of birth (available for Denmark, England, Norway, Portugal, Scotland), socioeconomic status/deprivation indicator (available for England, Norway, Scotland), comorbidity (available for Denmark, Norway), and international travel within 14 days before positive test (available for Luxembourg). Full details on the adjustment variables are available in Supplementary Table S1.

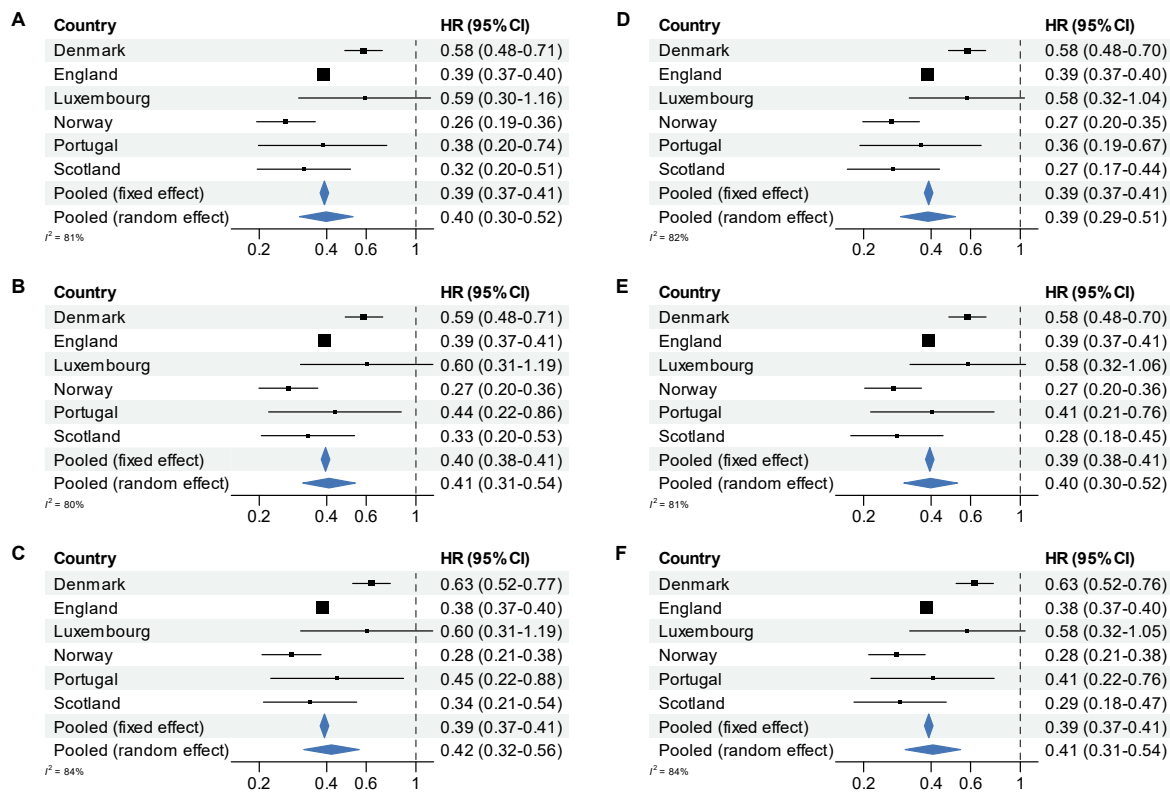

**Figure 3:** Hazard ratios of ICU admission (COVID-19-specific) for COVID-19 cases infected with the Omicron BA.1 versus Delta variants, by country and pooled across countries using a fixed effect and a random effect model.

(A). Adjusted for the required set of adjustment variables, using stratification for exact calendar date.
(B). Adjusted for the required and highly desired set of adjustment variables, using stratification for exact calendar date.
(C). Adjusted for the required, highly desired and desired set of adjustment variables, using stratification for exact calendar date.
(D). As (A), but using stratification for calendar week and regression adjustment for exact calendar date.
(E). As (B), but using stratification for calendar week and regression adjustment for exact calendar date.
(F). As (C), but using stratification for calendar week and regression adjustment for exact calendar date.

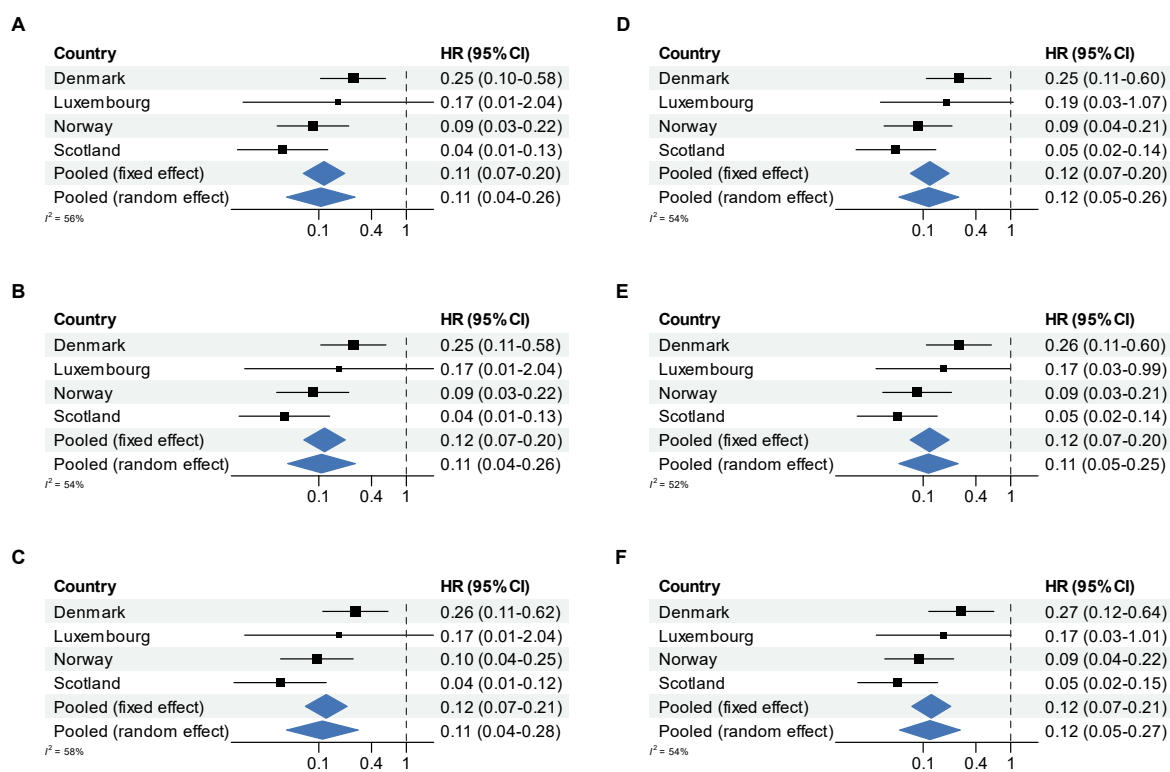

**Figure 4:** Hazard ratios of death (COVID-19-specific, where available, or otherwise due to any cause) for COVID-19 cases infected with the Omicron BA.1 versus Delta variants, by country and pooled across countries using a fixed effect and a random effect model.

(A). Adjusted for the required set of adjustment variables, using stratification for exact calendar date.
(B). Adjusted for the required and highly desired set of adjustment variables, using stratification for exact calendar date.
(C). Adjusted for the required, highly desired and desired set of adjustment variables, using stratification for exact calendar date.
(D). As (A), but using stratification for calendar week and regression adjustment for exact calendar date.
(E). As (B), but using stratification for calendar week and regression adjustment for exact calendar date.
(F). As (C), but using stratification for calendar week and regression adjustment for exact calendar date.

*Note:* In Portugal, no deaths were observed in Omicron cases; Portugal's results were therefore not included in the mortality analysis.

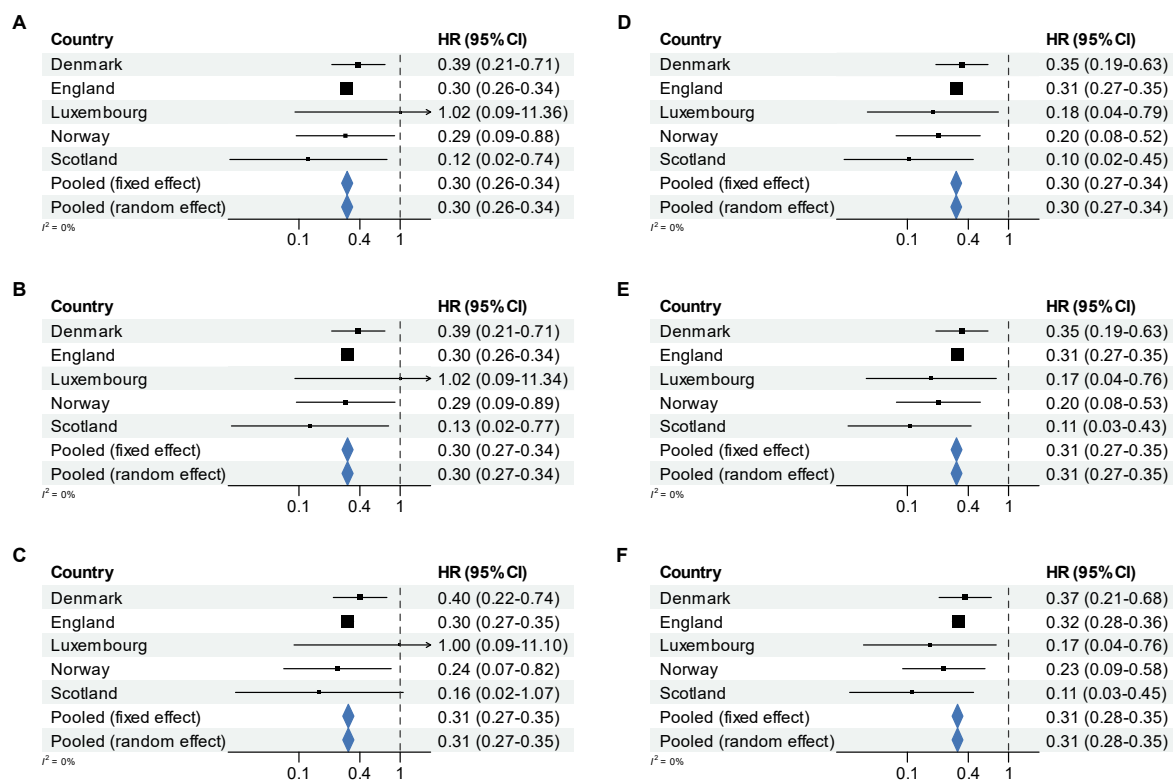

In subgroups by age (Figure 5; Supplementary Figures S8-S9), the HRs indicated greater reductions in risk between adult cases infected with Omicron versus Delta than those for children.

**Figure 5:** Hazard ratios of hospital admission (COVID-19-specific, where available, or otherwise due to any cause) for COVID-19 cases infected with the Omicron BA.1 versus Delta variants, as Figure 2F (adjusted for the required, highly desired and desired set of adjustment variables, using stratification for calendar week and regression adjustment for exact calendar date), by age group.

*Note:* In Portugal, no hospital admissions were observed in the 0-19 years age group; Portugal's results are therefore not included in that sub-analysis.

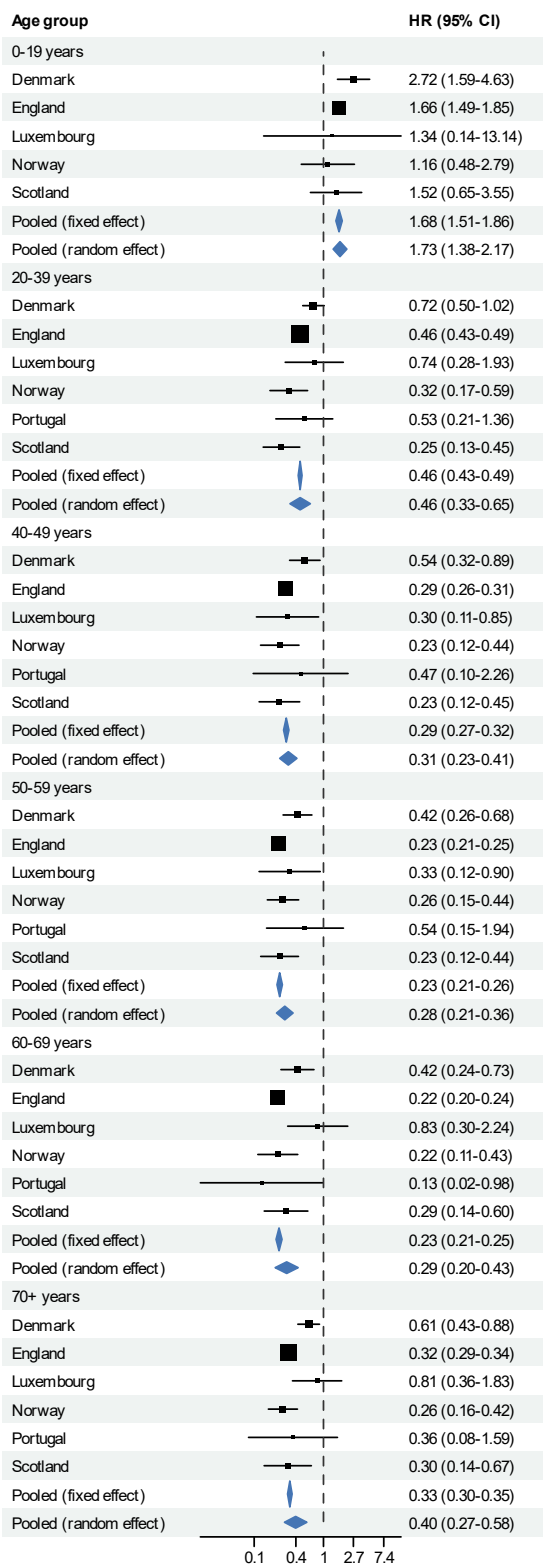

Split by vaccination status (Figure 6; Supplementary Figures S10-S11), the results indicated lower risks for Omicron compared to Delta cases in all subgroups. The estimated risks were similarly lower for Omicron compared to Delta for cases who were unvaccinated, and for cases who had received three vaccine doses. For those who had received one or two vaccine doses, the HRs were somewhat closer to 1.

**Figure 6:** Hazard ratios of hospital admission (COVID-19-specific, where available, or otherwise due to any cause) for COVID-19 cases infected with the Omicron BA.1 versus Delta variants, as Figure 2F (adjusted for the required, highly desired and desired set of adjustment variables, using stratification for calendar week and regression adjustment for exact calendar date), by vaccination status.

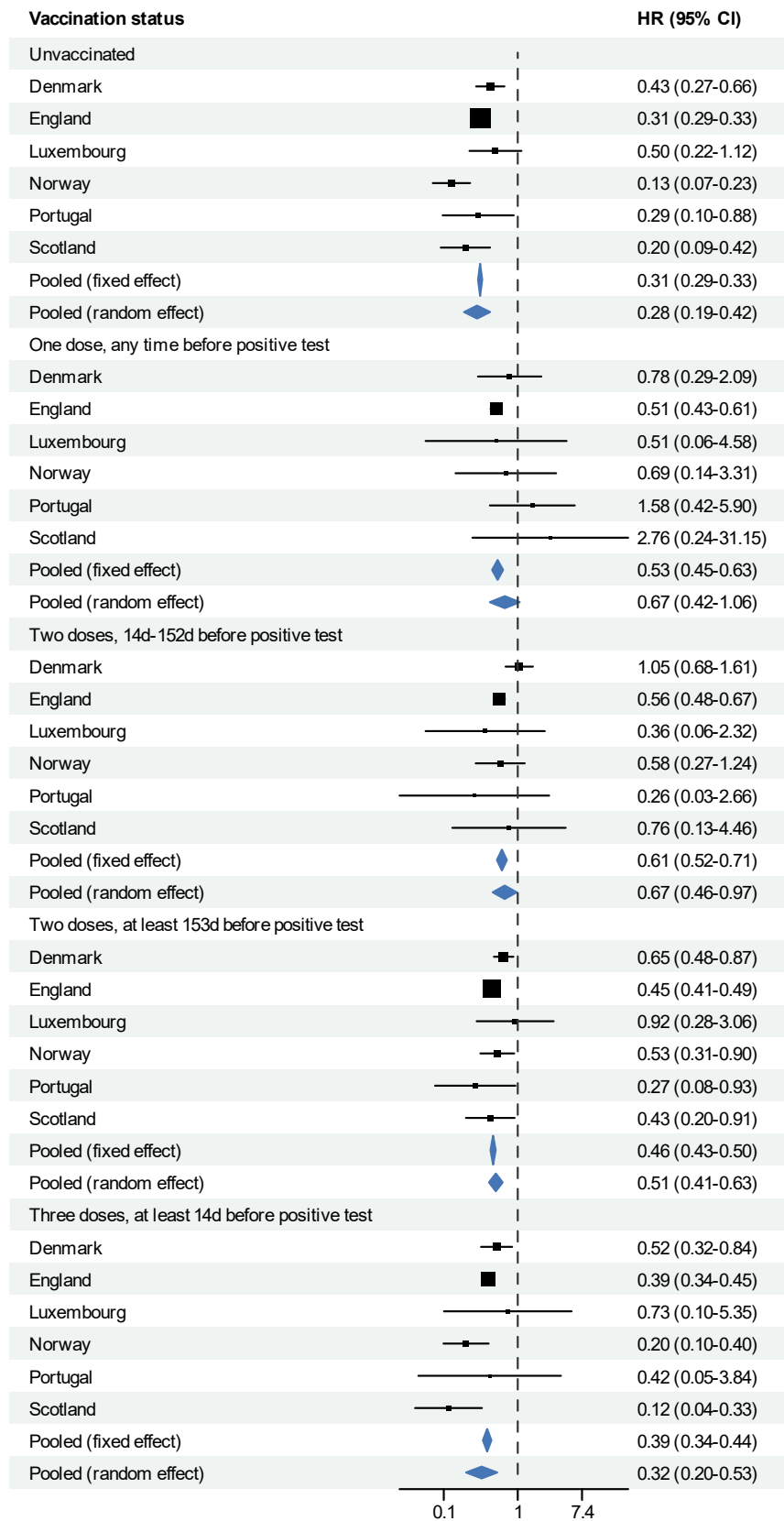

A sensitivity analysis to assess the impact of epidemic phase bias [24] suggested that the true HRs from all countries might be slightly lower than those estimated, i.e. indicating a slightly larger difference in risk between Omicron and Delta (Supplementary Figure S12). A sensitivity analysis that restricted to protocol-compliant Delta cases from Denmark whose variant had been reported as Delta (see Methods) found HRs closer to 1 than in the main analysis where all likely Delta cases were included (Supplementary Figure S13), potentially indicating that the variant classification had not been reported for a fraction of the hospitalised Delta cases.

The heterogeneity between the adjusted HR estimates was high for the hospital admission ($I^2$=84%, based on n=6 reporting countries) and ICU admission ($I^2$=54%, based on n=4 reporting countries) outcomes, but not for death ($I^2$=0%, based on n=5 reporting countries).

## Discussion

This is to our knowledge the first study to have combined data on the full national cohorts of COVID-19 cases available to public health agencies in several countries, to estimate the relative severity of those infected with different SARS-CoV-2 variants. This collaborative and standardised effort allows for the direct comparison and contrasting of the estimates from several countries, and has provided pooled estimates with even higher precision of the HRs of severe disease outcomes than that achievable based on analysis of single-country cohorts.

After adjustment for confounders, the pooled results indicate a 59% lower risk of hospital admission, 88% lower risk of ICU admission, and 69% lower risk of death for COVID-19 cases infected with Omicron BA.1 relative to those infected with the Delta variant. These lower risks are consistent with estimates reported from previous single-country analyses [1-9].

The results are also consistent with reports that the lower risk for cases with Omicron BA.1 compared to Delta was more pronounced in adults than in children [1,3,5,7,25], although some studies have reported conflicting data suggesting a similar reduction in risk for children and adults infected with Omicron [2,26]. For cases aged ≥20 years, the risk of hospital admission by age group was between 54% and 72% lower for Omicron BA.1 compared to Delta cases. For cases aged <20 years the results indicated higher hospitalisation risks for those infected with Omicron BA.1 than Delta. However, this should be considered in the context that absolute risks of severe COVID-19 outcomes are generally low for children regardless of infecting variant [1,3,5,7,25]. Moreover, previous analyses of those participant countries' data that suggested higher risks for children infected with Omicron BA.1 than Delta have suggested that differences in risk are considerably smaller when age groups are defined based on narrower age intervals than those considered here [1,3,5]. It has been suggested that the smaller differences in risks among children could be driven by a different symptom profile of the Omicron variant [3,27,28], which has a higher propensity to cause symptoms in the upper respiratory tract than in the lungs [29]. This might lead to a greater proportion of young children being admitted on cautionary grounds because of small airway obstruction.

By vaccination status, the results indicate approximately 70% lower risks of hospital admission for Omicron BA.1 compared to Delta cases who were unvaccinated, or who had received a booster dose, but smaller differences in risk between variants of 33% to 49% for cases who had received one or two vaccine doses. This is consistent with reports of lower vaccine effectiveness against hospitalisation for Omicron BA.1 breakthrough cases who had received only one or two vaccine doses than for corresponding Delta breakthrough cases [3,5,30,31]. Such lower vaccine effectiveness for Omicron BA.1 should be expected to lead to relatively smaller differences in risk between Omicron BA.1 and Delta in these vaccination subgroups, consistent with that observed.

There was high heterogeneity between the HR estimates from the individual countries for the hospital admission and ICU admission outcomes but not for the mortality outcome. Heterogeneity between national estimates may be expected, and this underlines the added value of comparing and pooling estimates generated via a common protocol to obtain more reliable results for public health decision making. The heterogeneity might in part reflect differences between countries in outcome definitions for the hospital and ICU admission outcomes. For example, some of the countries considered only COVID-19-related admissions whereas some countries included all admissions, some country-specific hospital admission definitions restricted to hospital attendances lasting at least one night whereas some country-specific definitions included any recorded admissions including those that lasted less than one day. By contrast, mortality definitions are unlikely to differ substantially between countries, consistent with the lack of heterogeneity in the HRs of this outcome. Moreover, the heterogeneity was lower between the countries' hospital admission HRs within subgroups by age or vaccination status, indicating that the heterogeneity may also in part be explained by differences in the distributions of these effect-modifying factors by country, due to e.g. differences in testing patterns by age group and vaccination schedules.

A previous multinational study reported on the severity of the Alpha, Beta and Gamma variants based on data from seven European countries reported to The European Surveillance System (TESSy) [16]. However, this past analysis included relatively small numbers of cases with classified variants (n per country, range 13 to 9,740) [16]. By contrast, the protocol and analysis proposed here enables investigators in individual countries in Europe and beyond to analyse their own entire national cohorts, and also allows for multinational collaborations to undertake comparative investigations and meta-analyses of the consistently analysed national datasets, as demonstrated in the Omicron BA.1 versus Delta pilot.

Currently, comprehensive community testing for COVID-19 has been reduced in many countries [19-22], which is likely to pose challenges for estimation of SARS-CoV-2 variant severity using cohort study designs as described here. The protocol is applicable to settings where data is available on a cohort of community cases, and is directly applicable in settings with continued community testing or in future pandemic scenarios with widespread community testing. However, assuming that COVID-19 cases who experience an outcome such as hospital admission can be identified and their infecting variant assessed, case-control type studies may be considered in the absence of wide-reaching community testing. Such a case-control study design could compare the prevalence of different variants between COVID-19 patients who are hospitalised and a suitably chosen comparison group of COVID-19 patients with less severe disease, e.g. cases identified through sentinel community surveillance. However, prevalence of a newly introduced variant generally changes with calendar time and hence the observed prevalence in a sample of COVID-19 cases will depend on their dates of infection. Therefore, care must be taken in the design to ensure that the two groups are comparable in terms of calendar time and the time from infection until testing. A preliminary analysis of data from England suggests that similar estimates of the relative risk as from the meta-analysis could be obtained by re-analysing the cohort data using a matched case-control design, provided that the mean time from positive test in the community to hospital admission is known, but that matching hospitalised cases on hospital admission date to community cases on community test date could lead to considerable bias. Further research should explore the feasibility of case-control type designs for surveillance of SARS-CoV-2 variant severity, in single countries and in international collaborations, including "right-sizing" of the optimal rate of community sampling that is required to enable reliable case-severity estimates.

Limitations of the study include a reliance on test-positive cases, which might make the results susceptible to selection bias if there were differences in testing patterns e.g. by calendar time and locality. However, all the included countries employed community testing for COVID-19 throughout the entire study period. Moreover, the use of stratification for calendar date and area in the data analysis, so that only cases identified under similar testing schedules are compared, should limit the impact of such bias. The protocol outlined several potential confounders as adjustment variables. Although all countries could adjust for the most likely confounders in the required and highly desired sets of adjustment variables, no country had the data to adjust for all adjustment variables classified as desired. However, the results from all countries indicated small differences between the HRs adjusted for any available variables additional to the adjustment variables classified as required, which might suggest that the magnitude of confounding due to the expanded set of adjustment variables is small. The outcome definitions were allowed to differ between countries. This was to accommodate the level of detail available for each country's data, to enable wider participation. As a result, the outcomes included both COVID-19-specific and all-cause outcomes. For the all-cause outcomes, incidental non-COVID-19-related events are likely to result in a small attenuation of the HRs compared to if COVID-19-specific outcomes had been available. Although this to our knowledge is the largest international collaborative study on Omicron BA.1 versus Delta severity to date, it included no more than six countries. One country, England, had considerably higher case numbers than the other included countries. Therefore, as expected, the HRs from England had great weight on the pooled estimates, particularly the fixed-effect estimates. This might limit the generalisability of the pooled estimates, especially in light of the high heterogeneity of the HRs between countries. However, this highlights the added value of contrasting the countries' estimates and exploring their heterogeneity.

Conclusions
This study has demonstrated the feasibility of conducting variant case-severity analyses in a multinational collaborative framework. The consistency in the direction of the HR estimates from the real-world analysis, and the high-precision pooled estimates further add to the evidence for the relatively lower severity of the Omicron BA.1 relative to the Delta variant. The proposed study protocol and accompanying statistical analysis code can facilitate future studies on relative severity by investigators in countries with community testing and available test, virus variant and severe outcome data, which will support public health risk assessments both for future SARS-CoV-2 virus variants and other similar infections.

## Ethical statement

Ethical approval for each national analysis was either granted by local ethics committees or not required according to national law. A full description is available in the Supplementary material (Supplement B: Supplementary text).

## Funding statement

This work was supported by the World Health Organization (WHO) Regional Office for Europe (TN, AMP); UK Research and Innovation (UKRI) Medical Research Council (MRC) (AMP: [Unit Programme number MC/UU/00002/11]); UKRI MRC/Department of Health and Social Care (DHSC) National Institute for Health and Care Research (NIHR) COVID-19 rapid response call (TN, AMP: [MC/PC/19074]).

The funders played no direct role in the study. The views expressed are those of the authors and not necessarily those of the authors' institutions, the NIHR or the DHSC. The author affiliated with the WHO is alone responsible for the views expressed in this publication and they do not necessarily represent the decisions or policies of the WHO.

## Conflicts of interest

All authors declare that they have no conflicts of interest.

## Authors' contributions

TN, RP and AMP conceived of and designed the study. PB, IBS, DB, JE, TGK, JMcMenamin, JMossong, HM, AP-S, PPL, JS, ST, LV, RW and JW contributed to the national collection and/or curation of COVID-19 data in the participating countries. TN and AMP wrote the first draft of the study protocol. All authors contributed to the final revised study protocol. TN wrote the standardised statistical analysis code. TN, IBS, JMossong, AP-S, JS and JW performed the statistical analyses of the countries' national datasets. TN performed the meta-analysis. TN wrote the first draft of the manuscript. All authors contributed to a revised draft, and approved the final version of the manuscript. RP organised the Joint Working Group and supervised the project.

# Supplement A

Study protocol

# PROTOCOL: Estimating relative case-severity risks by variant

## Introduction

During the COVID-19 pandemic of 2020-2022, several new variants of the SARS-CoV-2 virus have evolved that have varied in severity. Several European countries monitor the prevalence of newly evolved or introduced variants ("new" variants) in incident COVID-19 cases and their outcomes, but due to limited numbers these efforts may have only allowed detection of moderate to large differences in severity in individual countries. To enable timely assessment of the relative severity of new variants compared to previously prevalent variants ("old" variants), this protocol proposes (a) a standardised approach to quantifying the relative risks of severe outcomes by variant; and (b) a collaborative effort to pool the results from assessments of relative variant severity between nations, borrowing strength across countries through a meta-analytic approach.

## Study Design

National cohort studies of test-positive cases analysed by local investigators, according to a standardised analysis plan outlined below. Statistical summary results (such as relative risk estimates) are collated from each participating country to be synthesised using meta-analysis methods. No individual-level data are transferred outside the participating countries.

The aim of the study described in this protocol is to assess the strength of association between SARS-CoV-2 variant and one or more severity outcomes. The protocol assumes that these outcomes are censored time-to-event type outcomes that may be analysed using survival analysis methods. Such data are expected from participants in the proposed collaborative study. The outlined approach is, however, applicable for studies on severe outcomes of other data types, e.g. binary outcomes, ordinal outcomes, count outcomes, or quantitative outcomes.

## Setting

Community testing for SARS-CoV-2 infection in participating European countries.

## Participants and Eligibility Criteria

Individual-level data on test-positive COVID-19 cases with available data on SARS-CoV-2 variant, called through whole genome sequencing, genotyping or proxy tests such as assessment of S gene positivity of PCR test specimen. The inclusion calendar period should be chosen as a consecutive period of dates when cases of both variants under study are present. The choice of inclusion period is however left to the discretion of the national study groups.

If the data include cases whose positive specimen have only been assessed via proxy variant tests such as S gene positivity, it is recommended that the inclusion period is restricted to calendar dates when the positive (PPV) and negative predictive values (NPV) of the proxy test to distinguish variants are >90%. PPV is here defined as the probability that cases called as having the new variant with the proxy test do have the new variant based on sequencing-confirmed variant calls, and NPV is defined as the probability that cases called as having the old variant with the proxy test do have the old variant based on sequencing-confirmed variant calls. PPV and NPV generally depend on the prevalence of the new and old variant among new test-positive cases. The calendar-date-specific PPV and NPV can be estimated based on subgroups of cases whose variant has been called both through sequencing and the proxy test.

Exclusion criteria:

- No data on SARS-CoV-2 variant.
- Sequencing- or genotyping-confirmed SARS-CoV-2 variants other than those under study (e.g. rare variants without widespread transmission in the community).
- Data linkage errors (e.g. missing ID number, failure to link with data on outcome or required confounder variables).
- Missing data on at least one of the underlined required confounder variables.
- Vaccination patterns that may indicate immunosuppression:
  - ≥4 vaccination doses received, or
  - vaccination dose 3 received <80 days after dose 2.
- Other data inconsistencies (e.g. likely record errors). These should be specified by each national study group that reports cases excluded on these grounds.

Documentation of ethical approval for use of the individual-level data for research purposes must be provided by each participating national study group.

## Variables

### Exposure

- SARS-CoV-2 variant or variant sublineage, for example Omicron BA.2 vs Omicron BA.1. Determined based on (in order of precedence):
  a. Whole genome sequencing
  b. Genotyping (if applicable to distinguish lineages/sublineages under study)
  c. S gene positivity or other proxy methods, during calendar periods when calendar-time-specific positive and negative predictive values to distinguish the lineages/sublineages under study is >90%.

### Outcomes

In general, the protocol may be followed to study association with any severity outcomes for which there are individual-level data available.

For the proposed collaborative study, the primary outcomes are chosen to be events (hospitalisations, ICU admissions or deaths) due to any cause, and secondary outcomes are chosen to be COVID-19-specific events. The reason for this choice is twofold: (1) events that require no additional data to classify their cause are more likely to be available than data on COVID-19-specific events in all participating countries, and (2) it is a methodological challenge to classify if events are due to COVID-19 and so it would be difficult to standardise a classification that is possible to apply with the data available in different countries.

*Collaborative study, primary outcomes*
1. Any hospital attendance or admission (any cause; including emergency care attendance or admission through emergency care) within 0-14 days after positive test.
2. Hospital admission (any cause; including admission through emergency care) within 0-14 days after positive test.
3. Admission to intensive care (any cause) within 0-14 days after positive test.
4. Death (any cause) within 0-28 days after positive test.

*Collaborative study, secondary outcomes (if available)*
5. Any hospital attendance or admission (COVID-19-specific; including emergency care attendance or admission through emergency care) within 0-14 days after positive test.
6. Hospital admission (COVID-19-specific; including admission through emergency care) within 0-14 days after positive test.
7. Admission to intensive care (COVID-19-specific) within 0-14 days after positive test.
8. Death (COVID-19-specific) within 0-28 days after positive test.

The classification of which events are COVID-19-specific will not be standardised and is allowed to differ between countries. The classification from each national study group should be reported to the study coordinators. For example, COVID-19-specific events could be defined as: hospital admissions with COVID-19 specific ICD10 codes; or deaths with COVID-19 as one of the causes of death on the death registration.

For the death outcomes (outcomes 4 and 8), the event-specific follow-up time should be the time from specimen to death if within 28 days, or otherwise the time from specimen to the earliest of the censoring time points (1) the date of latest follow-up/data extraction or (2) 28 days after specimen date. For the hospitalisation/intensive care outcomes (outcomes 1-3 & 5-7), the event-specific follow-up time should be the time from specimen to the event, if within 14 days after specimen, or otherwise the time from specimen to the earliest of the censoring time points (1) date of death, (2) date of latest follow-up/data extraction or (3) 14 days after specimen date. Follow-up times of 0 days (e.g. events on the same day as positive test) should be reset to 0.5 days.

Model and confounder variables

The association between variant (or variant sublineage) and severity outcomes may be confounded by other variables by which the risk of the outcomes varies. In principle, the association should only be confounded to the extent that these variables are differentially associated with the variants/variant sublineages. It is therefore suggested that special care is taken to adjust for calendar date, area/region and vaccination status: when a new variant is replacing an old variant, variant prevalence often varies considerably by calendar time and locality. Calendar time and locality may also be associated with severity outcomes, e.g. through differences in time-and-place-specific healthcare burden or practice. Vaccination status may be associated with variant, if the studied variants have different propensities for breakthrough infections (Andrews et al., 2022), and also with the risk of the outcomes, since the available vaccines offer protection against severe disease in breakthrough cases (Collie et al., 2022).

In addition, several variables have been reported to be associated with risk of severe disease, such as age, sex, comorbidity and socioeconomic factors (Khawaja et al., 2020), although these may not in general be expected to be strongly associated with variant (or variant sublineage). Similarly, some variables such as recent international travel may be suspected to be associated with variant in countries where the new variant is (more recently) imported, but not necessarily strongly associated with the outcome(s). If available, adjustment for these variables may be carried out to rule out potential confounding.

For the collaborative study, two sets of adjustment variables are proposed: (1) one required, minimum set of adjustment variables for which format and categories should be standardised; and (2) one set of additional desired adjustment variables where formatting and categories are allowed to differ to a greater extent according to national setting. Data on all required variables are needed to participate in the collaborative study, whereas research groups may participate regardless of available data on the desired variables.

| Variable | Description | Data type | Categorisation (if applicable) |
|---|---|---|---|
| **Required** | | | |
| VOC | SARS-CoV-2 variant or variant sublineage. | Categorical variable. | ● [Old variant]<br>● [New variant] |
| DATE | Calendar date of positive test. Earliest positive test in the most recent infection episode. | Date variable. | |
| WEEK | ISO calendar week of positive test. | Categorical variable. | Defined based on the DATE variable. Categories are determined by the inclusion period. |
| AREA* | Area/region of residence. | Categorical variable. | Categories are allowed to differ according to national setting, but are recommended to be chosen so as to align with healthcare administrative areas/regions. |
| | | | For geographically small countries, a single region may be used for the entire country. |
| VACC | Vaccination status at date of positive test. | Categorical variable. | Preferred categorisation:<br>● Unvaccinated<br>● <28 days after first dose<br>● ≥28 days after first dose and <14 days after second dose, if any<br>● 14-152 days after second dose and <14 days after third dose, if any<br>● ≥153 days after second dose and <14 days after third dose, if any<br>● ≥14 days after third dose<br>If the vaccination data available do not allow for the above categorisation, the study groups are asked to contact the study coordinators with a description of the available vaccination data. |
| AGE | Age on date of positive test (in years). | Numeric variable. | |
| AGE10YR | Age group (10- or 20-year bands). | Categorical variable. | Defined based on the AGE variable.†<br>● 0-19 years<br>● 20-39 years<br>● 40-49 years<br>● 50-59 years<br>● 60-69 years<br>● ≥70 years |
| SEX | Biological sex. | Categorical variable. | ● Female<br>● Male |
| **Highly desired** | | | |
| REINF | Known reinfection, defined as ≥1 known past positive test >90 days before current infection episode. | Categorical variable. | ● No known past infection.<br>● Known past infection. |
| **Desired** | | | |
| ETHN* | Ethnic group, and/or country of birth. | Categorical variable(s). | Categories allowed to differ according to national setting. Include "Unknown" as a category for cases with missing data (if any). |
| SES* | Socioeconomic status/deprivation indicators. | Categorical variable(s). | Categories allowed to differ according to national setting. Include "Unknown" as a category for cases with missing data (if |

| | | | any). |
|---|---|---|---|
| COMORB* | Comorbidity. | Categorical variable(s). | Charlson Comorbidity Index (preferred), or otherwise allowed to differ according to national setting. Include "Unknown" as a category for cases with missing data (if any). |
| TRAVEL | International travel from any country within 14 days before testing positive. | Categorical variable. | ● No international travel within 14 days before testing positive.<br>● International travel within 14 days before testing positive.<br>Include "Unknown" as a category for cases with missing data (if any). |

\* Categories are allowed to differ according to national setting.

If there are potentially relevant data available on some of the above variables but the available data do not allow for the classification to be in the recommended format, the study groups are asked to please contact the study coordinators to discuss their applicability.

## Data sources / linkage
Case- and/or episode-based individual-level data record-linked (if necessary) to outcome, confounder and variant datasets.

## Statistical methods
To provide relative risk estimates that are adjusted for potential confounders, it is recommended that the investigators use conditional methods such as stratification for the confounders considered most likely to confound the association a priori. Stratification is a statistical method that is conceptually similar to matching, but where data from all potential matches are used instead of e.g. a fixed number of comparison cases with an old variant for each case with a new variant. Stratified Cox regression can be used where the outcome(s) are censored time-to-event type data, allowing for a different baseline hazard in each stratum. It is readily available in most standard statistical software. Similar methods are available for other types of outcome data. For example, conditional logistic regression may be used for binary outcomes, or conditional Poisson regression may be used for binary or count outcomes, to achieve similar stratification-based adjustment. The study coordinators will share standardised and fully documented R code for stratified Cox regression, and will advise on implementation.

## For the collaborative study

A summary of the number of cases included and excluded should be provided by each participating study, in a standardised format provided in the Appendix (Appendix Figure 1). This summary should provide sufficient information to calculate the proportion of all cases in the country during the inclusion period that were included in the analysis. The same summary figure may be used if the protocol is used in separate studies.

Descriptive frequencies of the distribution of the confounders by the cases' SARS-CoV-2 variant or variant sublineage should be provided from each participating study using the template provided in the Appendix (Appendix Table 1).

Stratified Cox proportional hazards regression should be used to estimate hazard ratios of new vs old variant lineage/sublineage, while adjusting for confounders and accounting for administrative censoring due to not all cases necessarily having a complete follow-up of 14 or 28 days.

Because of the potential association with both exposure and outcome for vaccination status, and since adjustment for calendar-period-and-locality-specific confounders such as healthcare burden is only through calendar date and area information, stratification-based methods should be used to ensure that the analysis is informed only by cases with the same vaccination status, from the same calendar periods and areas. Two alternative stratification approaches should be used: one that prioritises stricter control for exact calendar date (but which may not be feasible for smaller sample sizes); and one that prioritises somewhat higher precision while still matching for calendar week and adjusting for exact calendar date:

- Stratification for vaccination status, area and exact calendar date of specimen.
- Stratification for vaccination status, area and ISO calendar week of specimen, and including an interaction term between calendar week and linear calendar date.

### Primary analyses

The following primary models should be fitted. Note that missing data on any of the included <u>required</u> variables is an exclusion criterion.

**Model 1a.** The primary model should include the effects of:

Stratification for:
- DATE
- AREA
- VACC

Fixed effects regression terms for:
- VOC

- AGE10YR
- AGE10YR ✕ AGE (interaction term)
- SEX

**Model 1b.** A model with the same variables as in Model 1a, but with stratification for exact calendar date replaced by:

Stratification for:

- WEEK
- AREA
- VACC

Fixed effects regression terms for:

- VOC
- AGE10YR
- AGE10YR ✕ AGE (interaction term)
- SEX
- WEEK ✕ DATE (interaction term)

Secondary analyses

Secondarily, two more sets of models may be fitted if the investigators have access to data on additional confounders.

**Models 2a and 2b.** Models including all variables in Models 1a and 1b respectively, with additionally a fixed effects regression term for:

- REINF (if available).

**Models 3a and 3b.** Models including all variables in Models 1a and 1b respectively, with additionally fixed effects regression terms for all available variables out of:

- REINF (if available).
- ETHN (if available).
- SES (if available).
- COMORB (if available).
- TRAVEL (if available).

If only a subset of the above variables are available, those may be used instead in this secondary model.

4Subgroup analyses

- By age groups: Models should be fitted in the same format as Models 1a, 1b, 2a, 2b, 3a and 3b, but with an additional interaction term between variant and age group to estimate age-group-specific HRs.
    - AGE10YR ✕ VOC (interaction term).

- By vaccination status: Models should be fitted in the same format as Models 1a, 1b, 2a, 2b, 3a and 3b, but with an additional interaction term between variant and vaccination status to estimate vaccination-status-specific HRs.
    - VACC ✕ VOC (interaction term).

- By vaccination status and reinfection status (if data on reinfection status available): Models should be fitted in the same format as Models 2a, 2b, 3a and 3b, but with additional interaction terms between variant, vaccination status and reinfection status to estimate vaccination-and-reinfection-status-specific HRs.
    - VACC ✕ VOC + REINF ✕ VOC + VACC ✕ REINF ✕ VOC (interaction terms).

- By vaccination status and age group: Models should be fitted in the same format as Models 2a, 2b, 3a and 3b, but with additional interaction terms between variant, vaccination status and age group to estimate vaccination-status-and-age-group-specific HRs.
    - VACC ✕ VOC + AGE10YR ✕ VOC + VACC ✕ AGE10YR ✕ VOC (interaction terms).

Sensitivity analyses

The local investigators are recommended to investigate the sensitivity of the hazard ratios to the variant classification methods, particularly as sequencing may be biased towards more severe/hospitalised cases, although this potential bias is not expected to vary by variant. Furthermore, relative severity estimates may be affected by so-called epidemic phase bias (Seaman et al., 2022). Epidemic phase bias may occur when comparing two variants, if: (1) date of positive test is used to define inclusion period and/or is an adjustment variable; (2) the incidence of one variant is growing and the incidence of the other variant is declining, or where the incidence of one variant is growing faster than the incidence of the other variant; and (3) there is a correlation between the time from infection to positive test and disease severity. In such situations, the observed risk associated with the faster growing variant may be overestimated compared to the observed risk associated with the declining/slower growing variant (Seaman et al., 2022).

- Sensitivity to variant classification: repeat Models 1a and 1b, restricted to cases whose variant or variant sublineage was called by each of the used variant calling methods, e.g.

sequencing (if available), genotyping (if available), or proxy tests such as S gene positivity (if available).

- Epidemic phase bias: A sensitivity analysis to investigate the impact of epidemic phase bias under different assumed scenarios was recently proposed, that involves refitting models to a calendar date that is shifted for cases who experience severe outcomes, according to an assumed difference between the mean time from infection to positive test in cases who do not experience severe outcomes and the mean time from infection to positive test in cases who do experience severe outcomes (Seaman et al., 2022). It is recommended that the models are repeated under a range of scenarios with assumed differences ranging from 1 to 4 days. Code and detailed guidance can be provided by the study coordinators.

Further sensitivity analyses should be considered by local investigators, although they are not required to take part in the collaborative study synthesising estimates. These further assessments include:

- Sensitivity to adjustment strategy:
    - vary the number of adjustment variables and/or changing which variables are stratification or regression variables.

- Sensitivity to outcome definition(s):
    - modify the time period within which to consider severe outcomes (e.g. admissions or deaths within X days, varying X);
    - modify the criteria used to classify COVID-19-specific events.

## Outputs

For each of Models 1a, 1b, 2a, 2b, 3a and 3b and each sensitivity analysis, standardised code and guidance will be provided by the study coordinators to report the key effect size needed, i.e. the hazard ratio of each outcome for the new variant compared to the old variant.

For each of the subgroup analyses, the new variant:old variant hazard ratio of each outcome will be reported for each subgroup.

## Meta-analysis

After receiving the HR estimates and their associated standard errors from each country, the study coordinators will use fixed and random effects meta-analysis to combine the estimates from several countries into pooled HR estimates from each of the models (1a, 1b, 2a, 2b, 3a, 3b) and outcomes, and for each of the subgroup analysis models. The estimates from each country will be contrasted in forest plots and outliers will be investigated using funnel plots. The $I^2$ statistic will be used to formally quantify the heterogeneity of the estimates between countries.

## Strengths and limitations

Strengths of synthesising the results from several countries include the larger effective sample sizes that meta-analysis enables, which may lead to more rapid assessment of the risk associated with new variants. The analyses being run by separate national analysis teams ensures that local knowledge about the specifics of each country's data collection practices and other sources of bias are retained. Meta-analysis further allows for the results from different countries to be contrasted, which in turn will inform whether observed risk patterns with new variants are consistent between countries. Limitations include the reliance on pre-calculated estimates as opposed to if access and analysis of the individual-level data were centralised. However, the loss of precision of meta-analysis of country-level relative risk estimates compared to analysis of pooled individual-level data are likely acceptably small, and is a reasonable compromise that enables more rapid analysis of variant-specific risks than what international transfer of data would entail.

# APPENDIX Figure 1: Inclusion flow chart

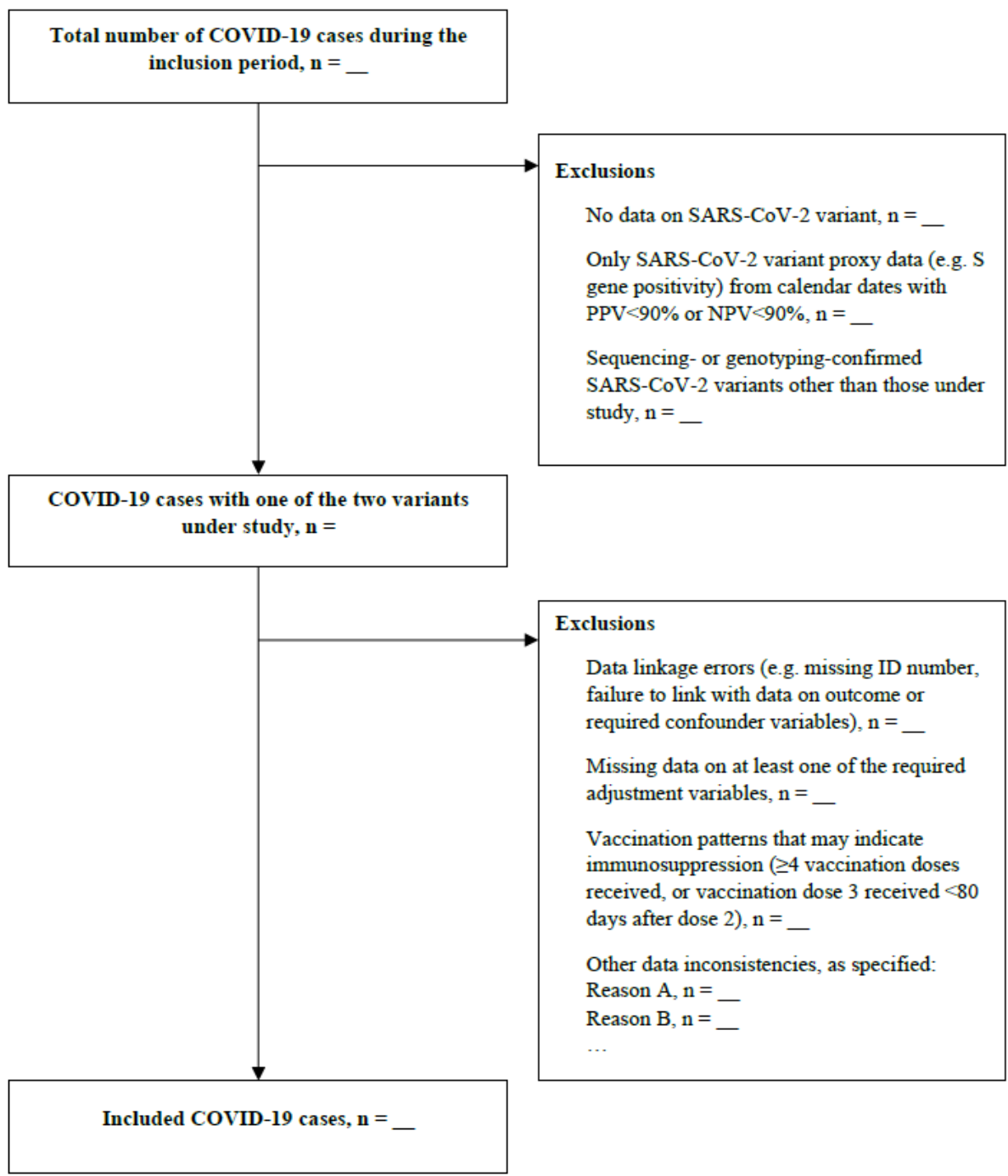

# APPENDIX Table 1: Characteristics

| Characteristic | Overall | [Old variant] | [New variant] |
|---|---|---|---|
| | n (%) | n (%) | n (%) |
| Total | | | |
| Age | | | |
|     0-19 | | | |
|     20-39 | | | |
|     40-49 | | | |
|     50-59 | | | |
|     60-69 | | | |
|     ≥70 | | | |
| Sex | | | |
|     Female | | | |
|     Male | | | |
| Calendar week of positive test | | | |
|     ISO week w | | | |
|     ISO week w+1 | | | |
|     ISO week w+2 | | | |
|     … | | | |
| Area/region of residence | | | |
|     Region A | | | |

| | | | | |
|---|---|---|---|---|
| | Region B | | | |
| | Region C | | | |
| | … | | | |
| Vaccination status at date of specimen | | | | |
| | Unvaccinated | | | |
| | <28 days after first dose | | | |
| | ≥28 days after first dose and <14 days after second dose | | | |
| | 14-152 days after second dose and <14 days after third dose | | | |
| | ≥153 days after second dose and <14 days after third dose | | | |
| | ≥14 days after third dose | | | |
| Reinfection status | | | | |
| | First infection episode | | | |
| | Reinfection episode | | | |
| Ethnicity/country of birth* | | | | |
| | Ethnicity A | | | |
| | Ethnicity B | | | |
| | Ethnicity C | | | |
| | … | | | |
| | Unknown | | | |
| Socioeconomic index/deprivation indicator(s)* | | | | |
| | Category A | | | |
| | Category B | | | |
| | Category C | | | |
| | … | | | |
| | Unknown | | | |
| Comorbidity* | | | | |
| | Comorbidity category A | | | |
| | Comorbidity category B | | | |
| | Comorbidity category C | | | |
| | … | | | |
| | Unknown | | | |
| International travel within 14 days before positive test | | | | |
| | No international travel | | | |
| | International travel | | | |
| | Unknown | | | |

* The number of variables and the categories to report for these factors is not standardised and is allowed to vary according to the data available in each local setting.

# Supplement B

Pilot to estimate Omicron BA.1 versus Delta variant severity

# Supplementary text: Country-level data descriptions

Denmark

### Data sources and linkage
We obtained data from the National Microbiology Database for all individuals tested with SARS-CoV-2 by RT-PCR in Denmark since March 1, 2020, and data from other national registers, available in the national COVID-19 surveillance system database at Statens Serum Institut (Copenhagen, Denmark), described in detail elsewhere [1]. Briefly, the surveillance system links individual-level information daily between national registers and databases by the use of the unique personal identification number of all Danish citizens, thereby centralising surveillance information from the National Patient Register (inpatient and outpatient diagnoses, admission and discharge dates), the National Vaccination Registry, the Civil Registration System (vital status and previous and current addresses), and variant-specific RT-PCR test results from the National Microbiology Database.

### Testing practices
Free-of-charge RT-PCR testing for close contacts of cases, those with mild COVID-19 symptoms, and asymptomatic individuals was provided by a centralized public COVID-19 test laboratory, Test Center Danmark (TCDK). All tests are provided as part of a universal tax-funded health-care system. Test timeslots are publicly available and can be booked online, and public tests sites are densely located across the country and open daily also for walk-in testing.

Individuals with symptoms suggestive of COVID-19 seen by a doctor (along with health-care personnel) were PCR tested in regional clinics and hospitals connected with the ten Danish departments of clinical microbiology (CMDs). CMDs serve both public and private hospitals, primary health clinics and long-term care facilities with suspected or ongoing outbreaks.

### Variant classification
Variants of cases analysed were determined using variant-specific RT-PCR testing. In the community track, an RT-PCR test using the L452 marker was developed and implemented by the Test Center Denmark (TCDK) by Dec 1, 2021. The estimated specificity was 99.99% based on retrospective analysis. In the health care track, eight of ten local Clinical Microbiology Departments (CMD) had similar omicron-specific variant-specific RT-PCR solutions set up and documented; the remaining two of ten local CMDs accounted for less than 7% of RT-PCR tests done by CMDs.

For the community track variant RT-PCR by TCDK, a confirmed omicron case was defined as a case testing positive with RT-PCR targeting the wild type sequence L452, as described elsewhere [2]. The delta variant is detected by the 452R substitution and was the only predominant variant in Denmark (>99%) in the period before occurrence of omicron; individuals testing positive for the 452R substitution were therefore considered to be infected with the delta VOC.

### Outcome data and definitions
A hospital admission was identified using three different data sources: 1) The National Patient Register which contains any contact to a Danish hospital longer than 12 h, 2) Twice daily hospital data (snapshots) in each region, which also includes admissions of duration less than 12 h on the condition that a bed is assigned to the patient, 3) All samples from hospitalised patients are confirmed as SARS-CoV-2 positive in the National Microbiology Database. COVID-19 specific hospital admission was defined by the main diagnosis of the hospitalisation having ICD10-codes DB342A, DB972A, DB972B, or DB948A. Information on ICU treatment was furthermore obtained from The National Patient Register. Hospital admission and ICU admission that occurred 0-14 days after positive test was considered.

Data on deaths were obtained from the Danish Civil Registration System and the Cause of Death Register. Deaths due to any cause were considered if they occurred within 0-28 days after positive test.

Ethical approval
The Danish study was performed as a surveillance study as part of the governmental institution Statens Serum Institut's advisory tasks for the Danish Ministry of Health. Statens Serum Institut's purpose is to prevent and fight the spread of infectious diseases in Denmark as specified in section 222 of the Danish Health Care Act. The need for ethics approval and informed consent is therefore deemed unnecessary according to national legislation, cf. implementing decree 2020-09-01, number 1338, about scientific regulatory procedure of health science research projects and health data scientific research projects. The presented data contains aggregated results and no personal data, and is therefore not subject to the European General Data Protection Regulation.

England

Data sources and linkage
COVID-19 is a notifiable disease in England, and all cases who test positive for the disease are required to be reported to the Second Generation Surveillance System (SGSS) at the UK Health Security Agency (UKHSA). This data resource includes baseline characteristics such as age, sex, ethnicity, region of residence, and index of multiple deprivation (an area-level classification of relative deprivation).

Here, data on all cases with positive test dates between 29 November and 30 December 2021 were extracted, and individually linked to variant classification results based on whole genome sequencing, provisional genotyping, or assessment of S gene positivity. These data were further linked to SARS-CoV-2 vaccination status through the National Immunisation Management Service; to hospitalisation data from two sources (Emergency Care Data Set and Secondary Uses Service); and to a mortality line list for COVID-19 cases maintained by UKHSA. All data were extracted and linked on 13 May 2022.

Testing practices
During the inclusion period, community testing was widely available, including cost-free lateral flow testing of asymptomatic individuals in the community, cost-free PCR-based community testing of symptomatic individuals, and targeted routine testing of healthcare staff, hospitalised individuals and international travellers.

Variant classification
During the study period, the variants of confirmed specimens were classified based on whole genome sequencing, provisional genotyping, or S gene positivity. Specimens were selected for whole genome sequencing or provisional genotyping based on geographically weighted population-level sampling of community cases, and targeted selection of hospitalised cases, hospital staff and recent international travellers. This variant classification was supplemented with S gene positivity data for specimens from England's community testing programme. Community test specimens were submitted for assessment with custom TaqPath assays at three laboratories that served the entire country depending on demand. The Delta variant is associated with S gene positivity and the Omicron BA.1 variant with S gene negativity, but S gene status is not specific to these two variants. Therefore, the inclusion period for the analysis was chosen to ensure a sufficient prevalence of both variants to ensure high predictive values when S gene positivity/negativity was used to classify variants as Delta or Omicron.

### Outcome data and definitions

Hospital admissions were extracted from two sources, the Emergency Care Data Set (ECDS) and the Secondary Uses Service (SUS). ECDS includes data on all attendances to emergency care departments in NHS hospitals, including those that result in admission, and is typically updated on a daily basis. SUS includes data on all hospital admissions to NHS hospitals, with reporting mandated at the time of discharge, and is updated monthly. Based on these two complementary data sources, hospitalisation events were classified as admissions within 14 days if the hospitalisation occurred within 0-14 days after positive test and (1) the time between admission and discharge was ≥1 day, (2) ECDS records indicated admission or transfer to another hospital department, or (3) the patient died in hospital on the same day as attendance.

Data on deaths in COVID-19 cases after positive test were collected by the UKHSA from four sources: (1) deaths in hospitals, (2) deaths notified to Health Protection Teams during outbreak management, (3) linkage of COVID-19 tests with death reports, and (4) death registrations where COVID-19 was mentioned on the death certificate. Here, deaths that occurred within 0-28 days due to any cause were considered.

### Ethical approval

This surveillance analysis was performed as part of UK Health Security Agency's (UKHSA) responsibility to monitor COVID-19 during the current pandemic. UKHSA has legal permission, provided by Regulation 3 of The Health Service (Control of Patient Information) Regulations 2002 to process confidential patient information under Sections 3(i) (a) to (c), 3(i)(d) (i) and (ii) and 3(iii) as part of its outbreak response activities. This study falls within the research activities approved by the UKHSA Research Ethics and Governance Group.

## Luxembourg

### Data sources and linkage

COVID-19 case were established from the contact tracing database (CARE+) which automatically receives all lab-confirmed cases. It also includes all antigenic results which are reported from pharmacies or health professionals.

Specific data collection systems were set up for monitoring hospital admissions and deaths from nursing homes and hospitals. Death outcomes were ascertained using death certificates. Vaccination status was obtained from a national vaccination COVID-19 database (MSVAC) set up for that particular purpose. Data bases were linked using a pseudonymised national personal identifier.

### Testing practices

PCR testing was widely available and did not change during the period, i.e. both symptomatic and asymptomatic cases could get tested. Testing was free following a medical prescription and for persons who self-declared a positive antigen test. There were twice weekly antigenic screening programmes in primary and secondary schools, as well as monthly screening in long term care facilities.

A twice weekly rapid antigen programme was in place in primary schools and once per week in secondary schools. All positive tests were in principle followed up by a PCR. Antigenic and PCR tests were also conducted in nursing homes when they had a cluster of cases.

Rapid antigen tests were available for purchase at relatively low cost in supermarkets and pharmacies. People were encouraged to test themselves prior to meeting up, e.g. for family get together sets.

### Variant classification

The national reference laboratory for acute respiratory infections receives SARS-CoV-2 positive samples (nasopharyngeal or oropharyngeal swabs analysed by RT-PCR) from the national network of laboratories. According to ECDC guidelines a representative sample of specimens chosen through random systemic selection is used for sequencing. Targeted sequencing was also conducted on hospitalised cases, breakthrough infections, or from large transmission clusters, e.g. in nursing homes. Overall, approximately 20% of confirmed cases were sequenced.

### Outcome data and definitions

Hospitals had to notify via a specific data collection tool, admissions which could potentially be related to a COVID-19: either suspected cases (e.g. in ICU based on scanner results), PCR test positive cases, and cases who were admitted for non-covid-19 reasons, but were positive at admission screening and who had to be isolated. ICU admissions included all hospitalisation events as above who went to ICU. Deaths were collected as any deaths that occurred within 28 days of the PCR positive test.

### Ethical approval

The current analysis is based on pseudonymized surveillance data collected by the Ministry of Health within the framework of the COVID-19 pandemic. The law on COVID-19 (https://legilux.public.lu/eli/etat/leg/loi/2020/07/17/a624/consolide/20210716) explicitly permits scientific research studies on the effects of vaccines against COVID-19 disease subject to prior pseudonymization.

## Norway

### Data sources and linkage

All data in this study for Norway came from the national emergency preparedness register for COVID-19, Beredt C19 (https://www.fhi.no/en/id/infectious-diseases/coronavirus/emergency-preparedness-register-for-covid-19/). Beredt C19 contains individual-level data from central health registries, national clinical registries and other national administrative registries. It covers all residents in Norway, and includes data on all laboratory-confirmed cases of COVID-19 in Norway, all hospitalizations among cases, deaths, and COVID-19 vaccinations. All databases were linked based on the national identity number.

### Testing practices

SARS-CoV-2 tests, both PCR and rapid antigen/lateral flow tests, were available free of charge for everyone during the study period, including those with mild or no symptoms, close contacts, and individuals in quarantine. Anyone who wished to be tested could get tested. Positive rapid antigen/lateral flow tests were recommended to be confirmed with PCR. Patients admitted to hospital were routinely tested for COVID-19. In addition, routine biweekly screening of school children with rapid antigen tests in areas with high transmission was recommended for secondary school students from late August 2021 and for primary school students from November 2021 until the end of January 2022. From 3 December 2021 until the start of February 2022, all travelers arriving to Norway were required to get tested.

### Variant classification

Data on virus variants came from the MSIS laboratory database (national laboratory database), which receives SARS-CoV-2 test results from all Norwegian microbiology laboratories. Variants were identified based on whole genome sequencing, Sanger partial S-gene sequencing or PCR screening targeting specific single nucleotide polymorphisms, insertions or deletions that reliably differentiate between Omicron and other variants. Details on SARS-CoV-2 variant surveillance and classification have been published by the Norwegian Institute of Public Health (in Norwegian): *https://www.fhi.no/nettpub/veileder-for-mikrobiologiske-laboratorieanalyser/covid-19/pavisning-og-overvakning-av-sars-cov-2-virusvarianter/*.

### Outcome data and definitions

We obtained data on hospitalisation following a positive SARS-CoV-2 test from the Norwegian Intensive Care and Pandemic Registry (NIPaR, part of Beredt C19) and data on COVID-19 related deaths from the Cause of Death Registry (DÅR, part of Beredt C19).

<u>COVID-19 hospitalisation</u>

In the study period, all Norwegian hospitals reported to NIPaR, and reporting was mandatory. Hospitals in Norway functioned within capacity during the study period, and criteria for hospitalisation and isolation for COVID-19 patients were consistent. During the study period, hospitals were encouraged to register new admissions within 24 hours. Reporting was timely. The median time from admission to registration in the period week 50 2021 – week 1 2022 was 0.9 days (interquartile range: 0.6 – 2.0 days. Hospitalisation was defined as hospital admission following a positive SARS-CoV-2 test, where COVID-19 was reported as the main cause of admission, based on a clinical assessment. Cases hospitalised with other or unknown main cause of admission were excluded from the study population in order to avoid bias. All admissions to hospital, regardless of length of stay, were included. Full details on the registration of patients hospitalised with COVID-19 are available here (in Norwegian): *https://helse-bergen.no/norsk-pandemiregister/registrering-i-norsk-pandemiregister-informasjon-til-ansatte.*

<u>ICU admission</u>

Patients are registered as ICU patients in NIPaR if they fulfil one of five categories:

- Length of stay over 24 hours in intensive care
- Require mechanical ventilation
- Are transferred between intensive care wards
- Persistent administration of vasoactive medication
- Length of stay under 24 hours, but passed away during stay in intensive care

Full details on the registration of ICU patients are available here (in Norwegian): *https://helse-bergen.no/norsk-intensivregister-nir/registrering-av-data-i-nir-kun-for-medlemmer.*

<u>COVID-19 deaths</u>

COVID-19 deaths are defined as deaths reported with COVID-19 as the main cause of death or contributing cause on the death certificate. More details are available here (in Norwegian):

https://www.fhi.no/sv/smittsomme-sykdommer/corona/dags--og-ukerapporter/sporsmal-og-svar-om-koronaovervaking-og-statistikk/.

Ethical approval

Ethical approval for this study in Norway was granted by Regional Committees for Medical Research Ethics - South East Norway, reference number 249509.

Portugal

Data sources and linkage

The study population was individuals eligible for vaccination (≥16 years old) diagnosed with SARS-CoV-2 infection/COVID-19 by nasopharyngeal swab tested with RT-PCR between December 1st and 29th 2021 notified through the laboratory service of the national surveillance system (SINAVE), in Portugal mainland. For this sample we excluded individuals with lateral flow tests. We included individuals with samples classified either by whole-genome sequencing or Spike Gene Target Failure (SGTF). Both symptomatic and asymptomatic individuals were included.

We obtained COVID-19 vaccination status through the electronic national vaccination register (VACINAS). Vaccination status was indexed at the date of COVID-19 diagnosis.

Information about age, sex and date of diagnosis, nationality and area of residency was routinely collected on surveillance system SINAVE and was extracted from it. Previous infection was defined as a PCR or rapid antigen SARS-CoV-2 notification for the same individual more than 90 days apart.

Testing practices

PCR testing is widely available and free of charge, but mainly to symptomatic individuals or to asymptomatic patients in healthcare settings. At the time of the study asymptomatic individuals had access to free of charge lateral flow tests. Most community pharmacies offered lateral flow tests to all the individuals regardless of their symptoms. All patients admitted to the hospital have a mandatory PCR test regardless of the symptoms. Patients in Long Term Care facilities might have routine testing.

Variant classification

We used information from a nationwide network group of laboratories (UNILABS, ABC and CVP) that performs RT-PCR tests for SARS-CoV-2 using Thermofisher TaqPath assay, targeting three regions of the SARS-CoV-2 genome: ORF1ab, N and S genes. Their samples are classified as BA.1 (no amplification) or B.1.617.2 (amplification), according to SGTF status. Only samples having both N and ORF1a positive signals and Ct values ≤30 were considered. Secondly, we use Whole-Genome Sequencing (WGS) data provided by the National Health Institute (INSA) that routinely performs sequencing on random samples notified to SINAVE. INSA follows the European Center for Disease and Control WGS sampling guidelines aiming at a 1% to 3% of total cases with WGS. Variants were classified first by whole genomic sequencing (WGS) and, if this information was unavailable, by detecting the S gene target failure (SGTF). The proportion of whole genomic sequencing (WGS) in the overall samples was 11% for Delta and 4% for Omicron BA.1, we used SGTF in the other samples to identify the variant.

Outcome data and definitions

We defined COVID-19 hospitalization as any admission to a public hospital in Portugal mainland within the 14 days following a positive sample SARS-CoV-2 collection, notified in the surveillance system, hence the last cases were followed until 12th of January. Admission data was obtained through the Central Hospital Morbidity Database and Integrated Hospital Information System. The Central Hospital Morbidity Database gathers data from all public hospitals in Portugal mainland, which account for most hospitals in Portugal and most hospitals admitting patients with COVID-19. Individuals with a SARS-CoV-2 diagnosis (by sample collection day) or notification after the admission date were excluded.

A COVID-19 death was defined as any record of death on the national Death Certificate Information System (SICO) with COVID-19 as the primary cause of death (ICD-10 *code U.071*) according to the WHO classification. SICO platform allows the issuance of a death certificate for each person who dies in Portugal [3]. A COVID-19 death was defined as any record of death on the national Death Certificate Information System (SICO) with COVID-19 as the primary cause of death (ICD-10 *code U.071*) according to the WHO classification. We followed the participants for 28 days after the sample collection day for the death outcome.

### Ethical approval
The genomic surveillance of SARS-CoV-2 in Portugal is regulated by the Assistant Secretary of State and Health Executive Order (Despacho n.º 331/2021 of January 11, 2021). The research on genomic epidemiology of SARS-COV-2 received the clearance of the Ethics Committee of INSA on March 30, 2021.

## Scotland

### Data sources and linkage
In Scotland, health records are indexed by the 10-digit Community Health Index number (CHI), a unique patient identifier allocated uniquely to individuals receiving health care.

All cases who test positive for COVID-19 the disease are reported to the ECOSS system at Public Health Scotland, recording baseline characteristics such as age, sex, ethnicity, region of residence, and index of multiple deprivation (an area-level classification of relative deprivation). Data on all cases with positive test dates between 01 May 2021 and 31 March 2021, with a Scottish postcode, were extracted (n = 1,167,065 cases), and individually linked to variant classification results based on whole genome sequencing, provisional genotyping, or assessment of S gene positivity. The majority of exclusions were a result of the variant not being typed, with n = 225,658 post linkage and initial exclusions. After linkage to SARS-CoV-2 vaccination status, and further exclusions for incomplete or inconsistent vaccination status, n = 215,586 cases.

147,872 cases with date of test between 10 October 2021 and 14 February 2022, the time period during which both Delta and Omicron typed cases were observed, were included in the analysis.

### Testing practices
During the inclusion period, community testing was widely available, including cost-free lateral flow testing of asymptomatic individuals in the community, cost-free PCR-based community testing of symptomatic individuals, and targeted routine testing of healthcare staff, hospitalised individuals and international travellers.

### Variant classification
Approximately 10% of all positive COVID-19 tests were sequenced across Scotland during the study period. The variants of confirmed specimens were classified based on whole genome sequencing, allele specific PCR testing, or S gene positivity. Specimens were selected for whole genome sequencing or allele specific PCR testing based on geographically weighted population-level sampling of community cases, and targeted selection of recent international travellers, fully vaccinated cases, immunocompromised patients, and hospitalised patients. This variant classification was supplemented with S gene positivity data for specimens from Scotland's community testing programme, which was available for approximately 90% of PCR positive results.

The Delta variant is associated with S gene positivity and the Omicron BA.1 variant with S gene negativity, but S gene status is not specific to these two variants. Therefore, the inclusion period for the analysis was chosen to ensure a sufficient prevalence of both variants to ensure high predictive values when S gene positivity/negativity was used to call variants as Delta or Omicron.

### Outcome data and definitions

Hospital admissions were identified from the General Acute Inpatient and Day Case - Scottish Morbidity Record (SMR01). This is a fully validated hospital admissions dataset created after a patient has been discharged from hospital and contains CHI, patient-level demographic details (treatment location, age, gender *etc*), completed ICD-10 codes and several other variables related to a patient's stay in hospital. There is a time lag of around 3 months before full completion and availability. Patients with long hospital stays may not be included, as both an admission and discharge date are required to create a record.

A relevant hospital admission was defined as a hospital admission within 14 days of a positive COVID-19 test. A COVID-19 related hospital admission was defined as a hospital admission within 14 days of a positive COVID-19 test with U07.1 (laboratory confirmed COVID-19) or U07.2 (clinical suspicion of COVID-19) in the primary diagnosis field.

ICU admissions were identified from the Scottish Intensive Care Society Audit Group (SICSAG) dataset that documents all adult Intensive Care Unit (ICU) and Higher Dependency Unit (HDU) admissions in Scotland. A relevant ICU admission was defined as an ICU admission within 14 days of a positive COVID-19 test. A COVID-19 related ICU admission was defined as a patient aged over 16 years who had been admitted to ICU within 14 days of a positive COVID-19 test, if U07.1 or U07.2 was entered as the patient's primary diagnosis for admission.

Data on deaths were obtained from National Records Scotland (NRS), an electronic record of all deaths registered within Scotland. This contains data on cause, time and place of death. Any cause mortality was defined as any death within 28 days of a positive COVID-19 test. COVID-19 specific mortality was defined as any death within 28 days of a positive COVID-19 test with U07.1 or U07.2 included within their cause of death codes (at any position).

### Ethical approval

This surveillance analysis was carried out by Public Health Scotland as part of its responsibility for the surveillance of respiratory pathogens in the population, and in accordance with UK GDPR Article 9 (2) i – the processing of personal data for the performance of a task for reasons of public interest in the area of public health.

**Supplementary Figure S1.** Inclusion summary flowchart for Denmark.

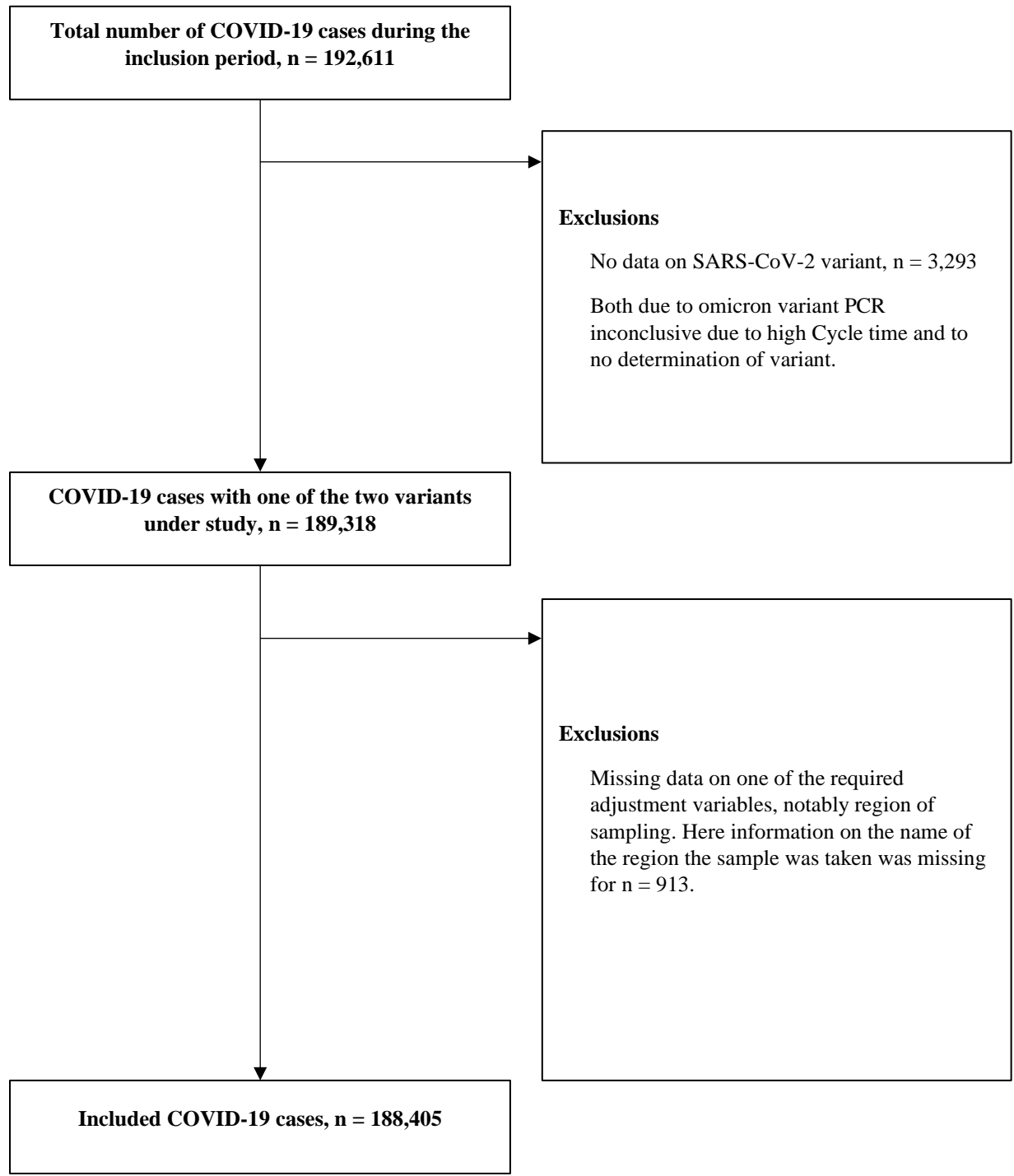

**Supplementary Figure S2.** Inclusion summary flowchart for England.

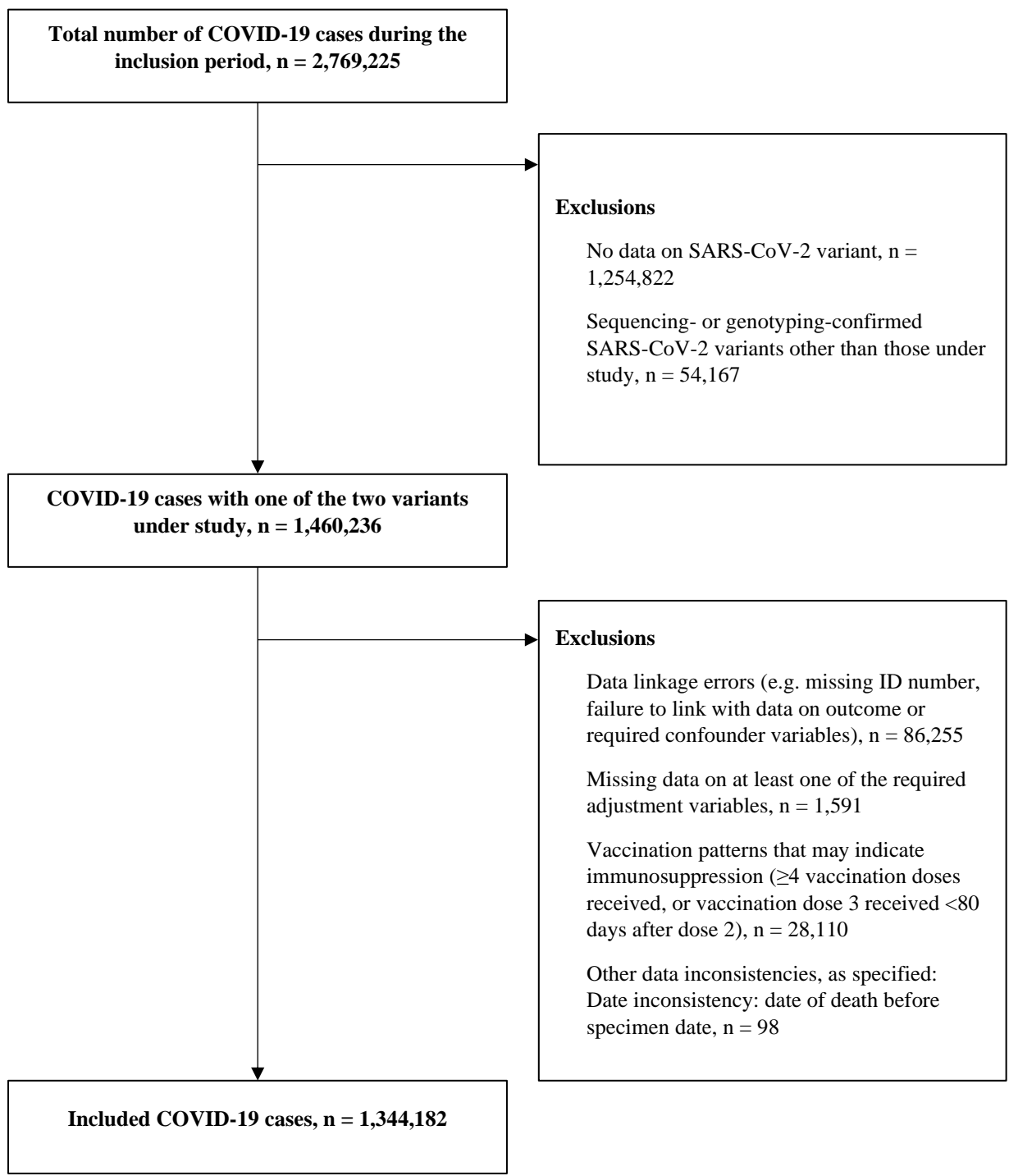

12**Supplementary Figure S3.** Inclusion summary flowchart for Luxembourg.

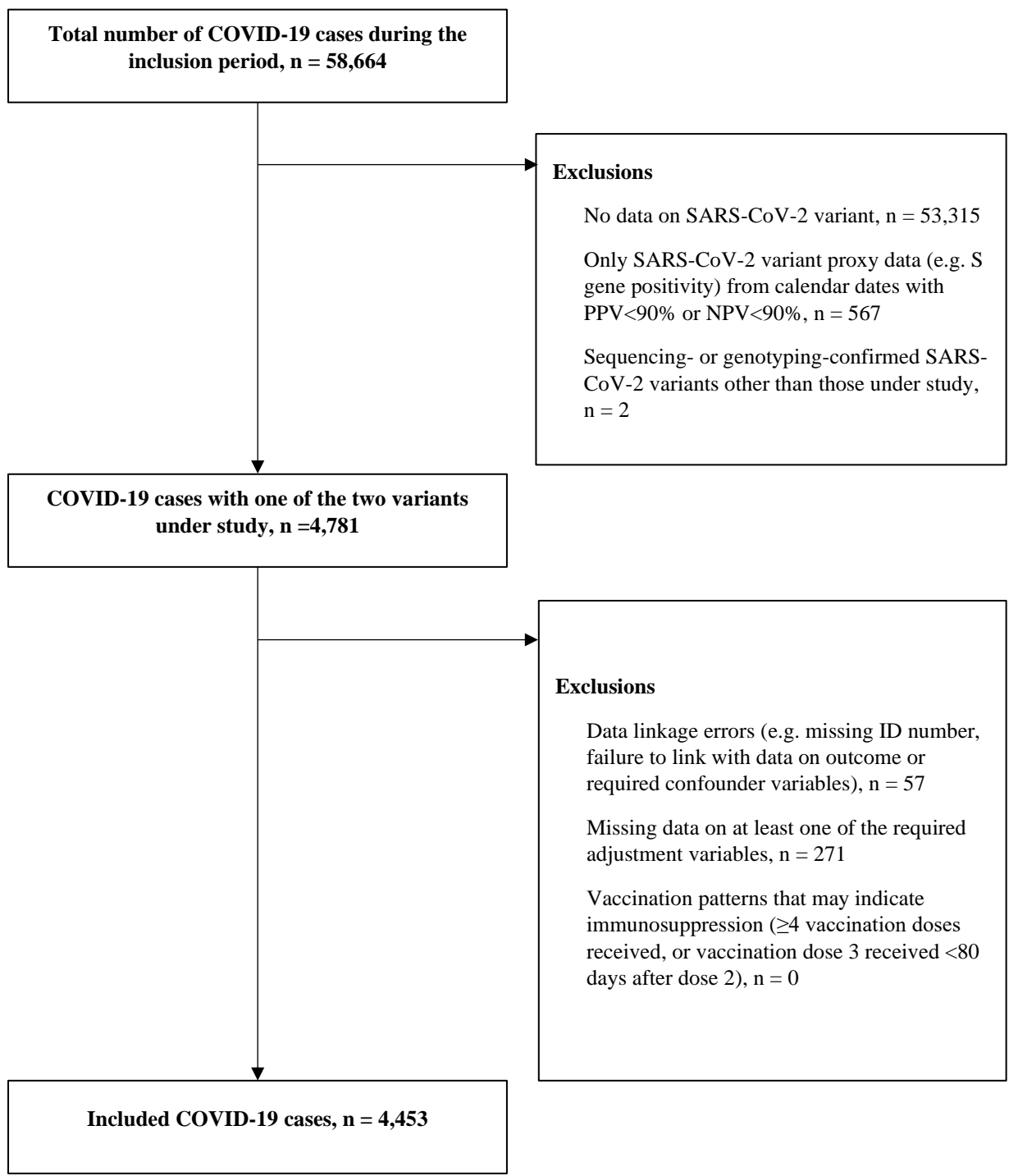

**Supplementary Figure S4.** Inclusion summary flowchart for Norway.

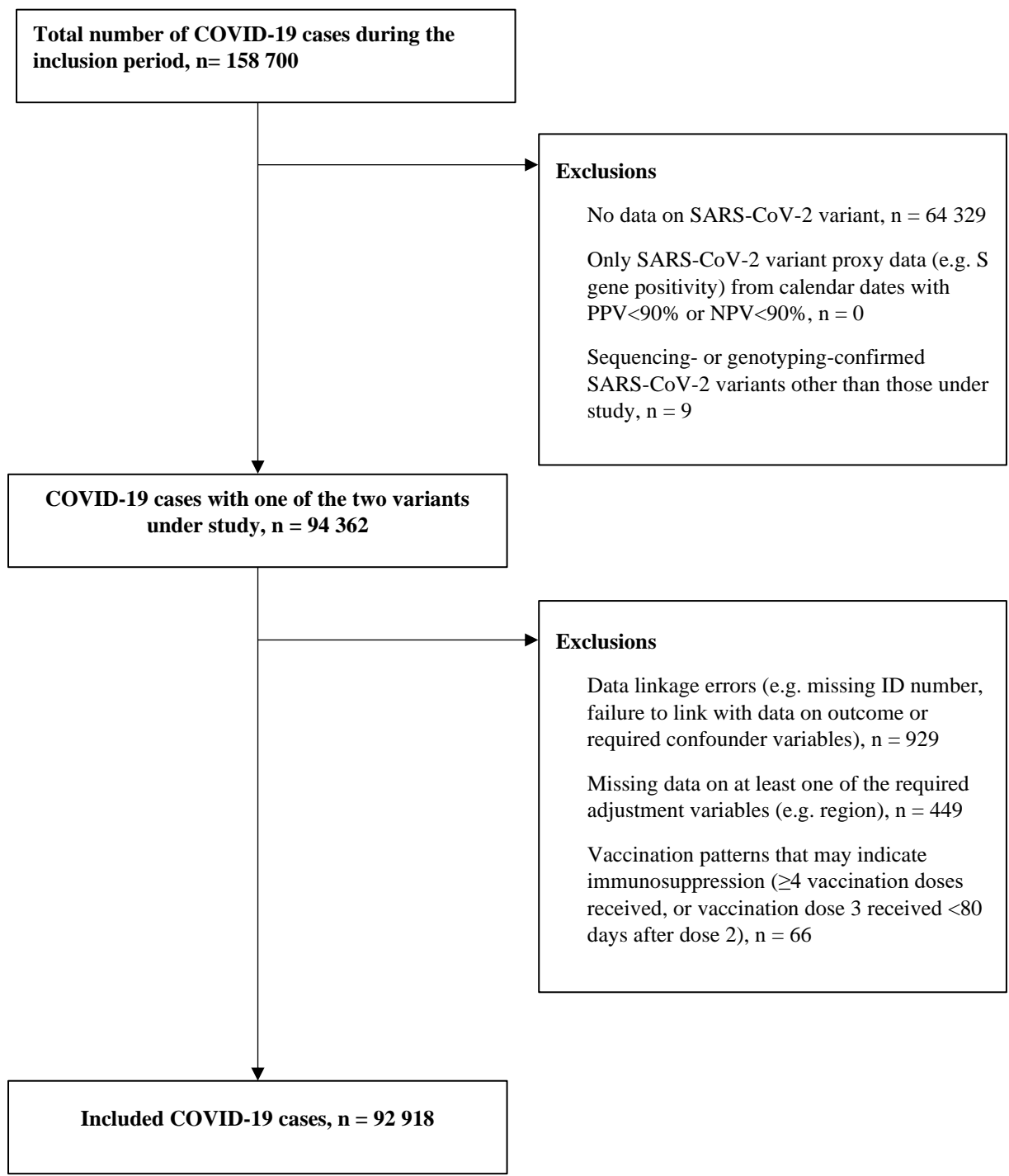

**Supplementary Figure S5.** Inclusion summary flowchart for Portugal.

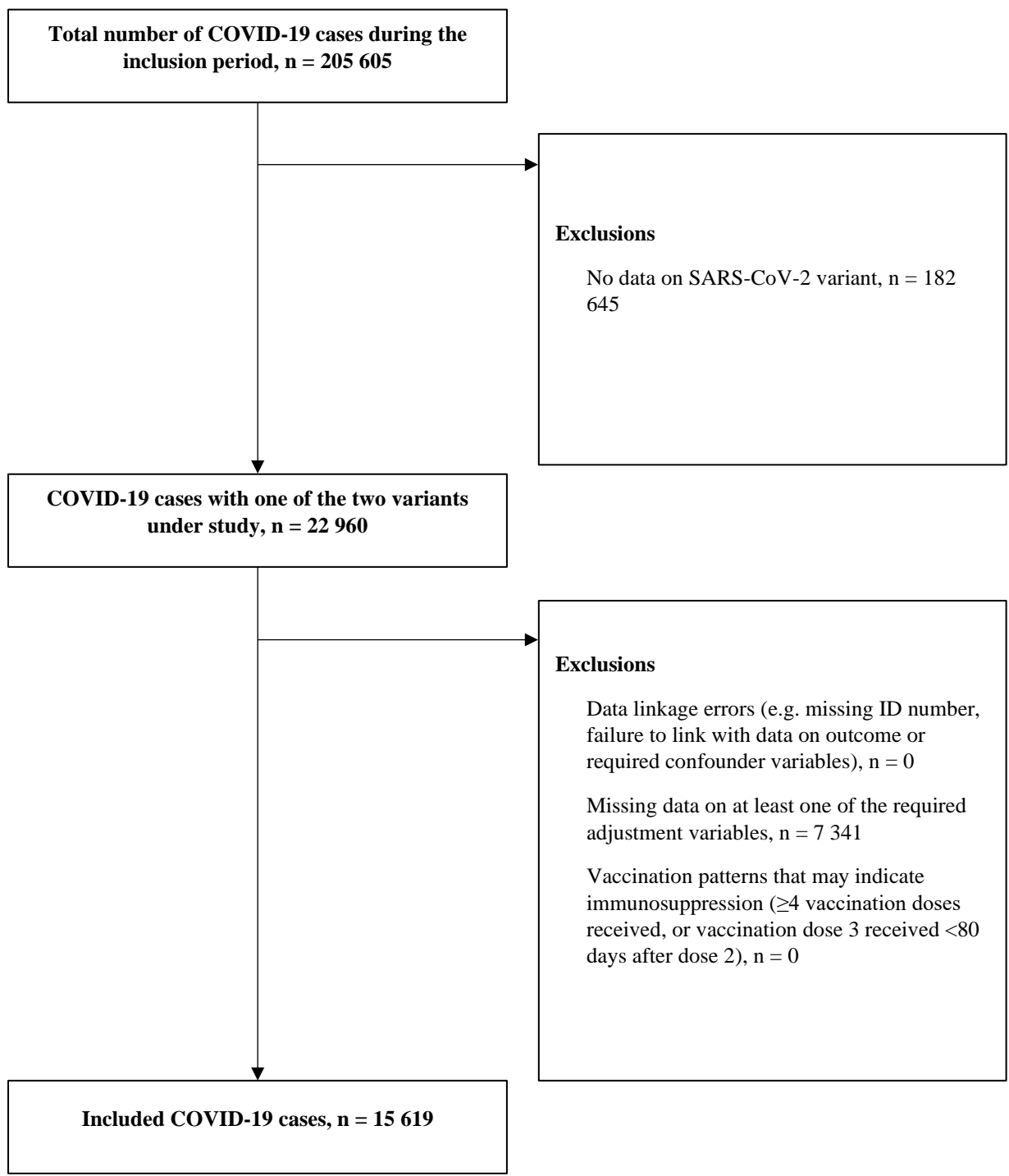

**Supplementary Figure S6.** Inclusion summary flowchart for Scotland.

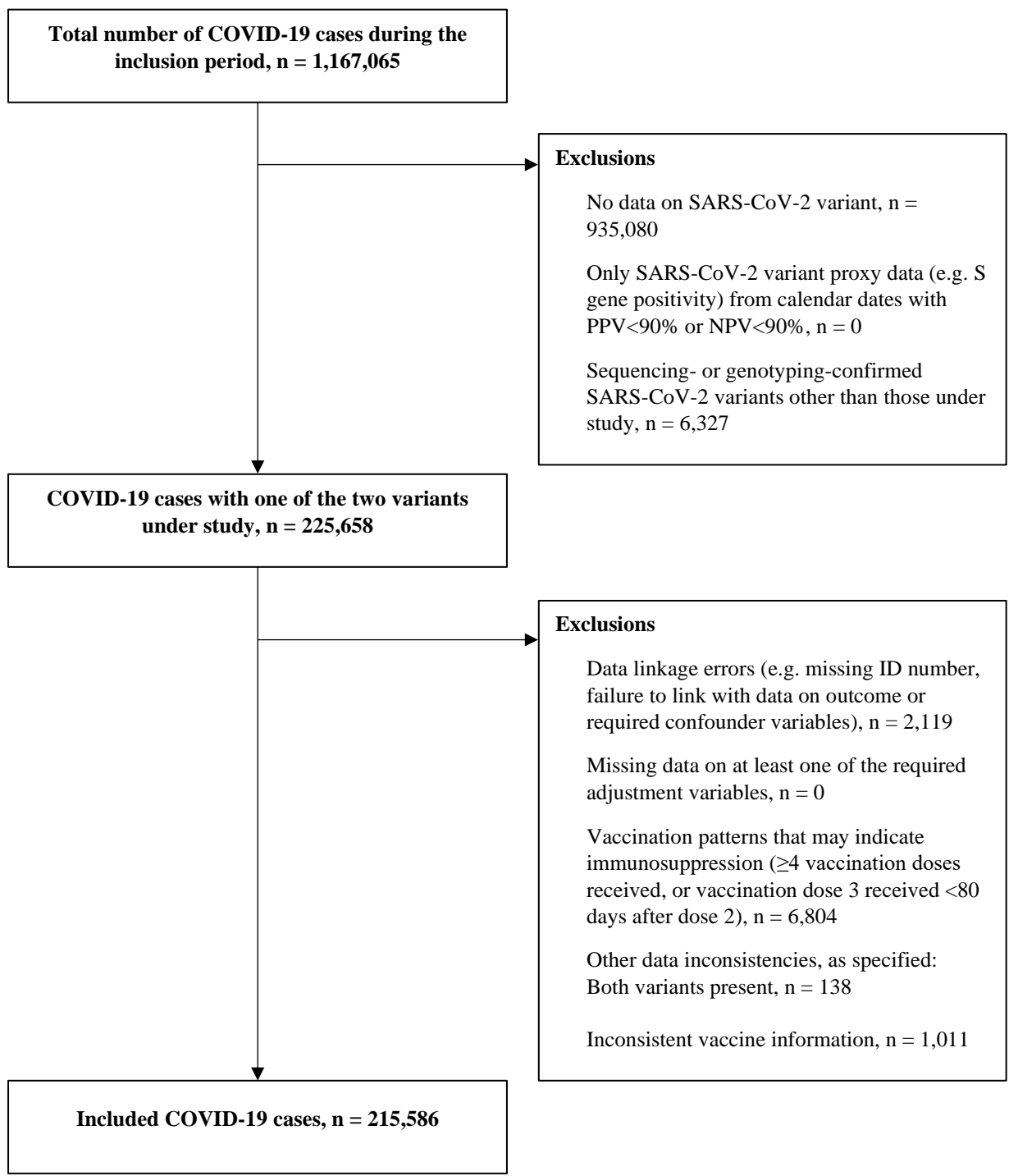

**Supplementary Table S1.** Summary of available outcomes and adjustment variables, by country.

| Category | Variable | Country | | | | | |
|---|---|---|---|---|---|---|---|
| | | Denmark | England | Luxembourg | Norway | Portugal | Scotland |
| **Outcome** | Hospital admission within 14 days after positive test | Available: Hospital admission due to COVID-19 (according to recorded ICD-10 codes) reported through three data sources, including hospital visits lasting longer than 12 hours, or lasting shorter than 12 hours but where a bed was assigned to the patient. | Available: Any hospital admission as recorded in two data sources covering admissions via emergency care or any hospital department where the time between admission and discharge was ≥1 day. | Available: Hospital admission for reasons potentially related to COVID-19 as determined and reported by hospitals. | Available: Hospital admission due to COVID-19 as reported to the national pandemic registry to which reporting was mandatory for hospitals during the study period, including cases who tested positive before admission or cases who tested positive during an ongoing hospitalisation. | Available: Any hospital admission as reported to a central hospital database. | Available: Hospital admissions due to COVID-19 (according to recorded ICD-10 codes). |
| | ICU admission within 14 days after positive test | Available: Hospital admission as defined above, to an ICU ward (as determined by hospital department coding). | -- | Available: Hospital admission as defined above, to an ICU ward. | Available: Admission to an ICU ward as recorded in the above pandemic registry, which fulfilled one of five criteria: (1) length of stay >24 hours in intensive care, (2) required mechanical ventilation, (3) transferred between intensive care wards, (4) persistent administration of vasoactive medication, or (5) length of stay under 24 hours but passed away during stay in intensive care. | Available: ICU admission as reported to the above central hospital database. | Available: Recorded ICU admission during hospital stay due to COVID-19 (as defined above). |
| | Death within 28 days after positive test | Available: Death as recorded in the population Civil Registration System and the Cause of Death Register. | Available: Deaths collected through hospital report and linkage with health records and death registrations. | Available: Death as recorded in population register with COVID-19 as primary cause of death. | Available: Death as recorded in population register, with COVID-19 as primary or contributing cause of death on the death certificate. | Available: Death as recorded in population register, with COVID-19 as primary cause of death on the death certificate. | Available: Death as recorded in population register, with COVID-19 as primary or contributing cause of death on the death certificate. |
| **Required adjustment variable** | Sex | Available with protocol-requested categories | Available with protocol-requested categories | Available with protocol-requested categories | Available with protocol-requested categories | Available with protocol-requested categories | Available with protocol-requested categories |
| | Age at diagnosis | Available (exact age) | Available (exact age) | Available (exact age) | Available (exact age) | Available (exact age) | Available (exact age) |
| | Calendar date of positive test | Available (exact date) | Available (exact date) | Available (exact date) | Available (exact date) | Available (exact date) | Available (exact date) |
| | Area of residence | Available: Copenhagen, Midtjylland, Nordjylland, Sjælland, Syddanmark | Available: East of England, London, Midlands, North East and Yorkshire, North West, South East, South West | Available: Luxembourg (whole country) | Available: East Norway, Oslo-wide area, South and West Norway, Mid-Norway, North Norway | Available: Algarve & Alentejo, Centro, Lisboa e Vale do Tejo, Norte | Available: Scotland (whole country) |
| | Vaccination status | Available with protocol-requested categories | Available with protocol-requested categories | Available with protocol-requested categories | Available with protocol-requested categories | Available with protocol-requested categories | Available with protocol-requested categories |
| **Highly desired adjustment variable** | Reinfection status | Available with protocol-requested categories | Available with protocol-requested categories | Available with protocol-requested categories | Available with protocol-requested categories | Available with protocol-requested categories | Available with protocol-requested categories |
| **Desired adjustment variable** | Ethnicity/country of birth | Available: Danish, second-generation immigrant, first-generation immigrant | Available: White, Asian, Black, Mixed/Other, Unknown | -- | Available: Born in Norway, Not born in Norway, Unknown | Available: Portuguese, Other | Available: White, Asian, Black, Mixed/Multiple, Other, Unknown |
| | Socioeconomic status/deprivation indicator | -- | Available: Index of multiple deprivation decile (10 categories) | -- | Available: Not overcrowded home, Overcrowded home, Overcrowding status unknown | -- | Available: Index of multiple deprivation quintile (5 categories) |

| | | | | | | |
|---|---|---|---|---|---|---|
| Comorbidity | Available: Number of comorbidities in the past 5 years (0 vs ≥1). The comorbidities include diabetes, adiposity, cancer, neurological diseases, nephrological diseases, haematological diseases, cardiac disease, respiratory diseases, immunological diseases, and other comorbid diseases. | -- | -- | Available: No underlying comorbidities, Medium-risk comorbidity, High-risk comorbidity [4] | -- | -- |
| International travel within 14 days before positive test | -- | -- | Available: No international travel within 14 days, International travel within 14 days | -- | -- | -- |

**Supplementary Table S2.** Descriptive outcome frequencies for included COVID-19 Delta or Omicron cases, by country

| Country | Outcome | n/N (%) Delta | Omicron |
|---|---|---|---|
| Denmark | Hospital admission (COVID-19-specific) | 1656/149689 (1.1%) | 143/38716 (0.37%) |
| | Hospital admission (any cause) | 2344/149689 (1.6%) | 258/38716 (0.67%) |
| | ICU admission (any cause) | 185/149689 (0.12%) | 6/38716 (0.015%) |
| | Death (any cause) | 264/149689 (0.18%) | 13/38716 (0.034%) |
| England | Hospital admission (any cause) | 8444/429231 (2.0%) | 8364/914951 (0.91%) |
| | Death (any cause) | 1342/429231 (0.31%) | 990/914951 (0.11%) |
| Luxembourg | Hospital admission (COVID-19-specific) | 70/1286 (5.4%) | 86/3167 (2.7%) |
| | ICU admission (COVID-19-specific) | 10/1286 (0.78%) | 4/3167 (0.13%) |
| | Death (COVID-19-specific) | 10/1286 (0.78%) | 14/3167 (0.44%) |
| Norway | Hospital admission (COVID-19-specific) | 520/52091 (1.0%) | 90/40827 (0.22%) |
| | ICU admission (COVID-19-specific) | 151/52091 (0.29%) | 10/40827 (0.024%) |
| | Death (COVID-19-specific) | 105/52091 (0.20%) | 17/40827 (0.042%) |
| Portugal | Hospital admission (any cause) | 139/9150 (1.5%) | 15/6469 (0.23%) |
| | Death (any cause) | 24/9150 (0.26%) | 0/6469 (0%) |
| Scotland | Hospital admission (COVID-19-specific) | 406/35534 (1.1%) | 117/47744 (0.25%) |
| | Hospital admission (any cause) | 510/35534 (1.4%) | 265/47744 (0.56%) |
| | ICU admission (COVID-19-specific) | 106/35534 (0.30%) | 9/47744 (0.019%) |
| | ICU admission (any cause) | 116/35534 (0.33%) | 17/47744 (0.036%) |
| | Death (COVID-19-specific) | 70/35534 (0.20%) | 9/47744 (0.019%) |
| | Death (any cause) | 72/35534 (0.20%) | 11/47744 (0.023%) |

**Supplementary Table S3.** Descriptive characteristics of included Delta or Omicron cases, by country

| Country | Category | Variable | Delta | Omicron | Overall |
|---|---|---|---|---|---|
| Denmark | | | (N=149689) | (N=38716) | (N=188405) |
| | Required adjustment variable | Sex | | | |
| | | Female | 75363 (50.3%) | 19450 (50.2%) | 94813 (50.3%) |
| | | Male | 74326 (49.7%) | 19266 (49.8%) | 93592 (49.7%) |
| | | Age (years) | | | |
| | | Mean (SD) | 32.9 (20.9) | 33.4 (16.7) | 33.0 (20.1) |
| | | Median (Q1-Q3) | 32.0 (12.0-49.0) | 29.0 (21.0-46.0) | 31.0 (14.0-48.0) |
| | | 0-19 | 53530 (35.8%) | 8915 (23.0%) | 62445 (33.1%) |
| | | 20-39 | 38700 (25.9%) | 16980 (43.9%) | 55680 (29.6%) |
| | | 40-49 | 23584 (15.8%) | 5519 (14.3%) | 29103 (15.4%) |
| | | 50-59 | 17075 (11.4%) | 4706 (12.2%) | 21781 (11.6%) |
| | | 60-69 | 9895 (6.6%) | 1846 (4.8%) | 11741 (6.2%) |
| | | ≥70 | 6905 (4.6%) | 750 (1.9%) | 7655 (4.1%) |
| | | Calendar week | | | |
| | | ISO week 46, 2021 | <3668 (2.4%) | <5 (0.0%) | 3668 (1.9%) |
| | | ISO week 47, 2021 | <28750 (19.2%) | <25 (0.1%) | 28766 (15.3%) |
| | | ISO week 48, 2021 | 33289 (22.2%) | 521 (1.3%) | 33810 (17.9%) |
| | | ISO week 49, 2021 | 41842 (28.0%) | 5951 (15.4%) | 47793 (25.4%) |
| | | ISO week 50, 2021 | 42147 (28.2%) | 32221 (83.2%) | 74368 (39.5%) |
| | | Area of residence | | | |
| | | Hovedstaden | 65926 (44.0%) | 23573 (60.9%) | 89499 (47.5%) |
| | | Midtjylland | 25277 (16.9%) | 6790 (17.5%) | 32067 (17.0%) |
| | | Nordjylland | 10603 (7.1%) | 1816 (4.7%) | 12419 (6.6%) |
| | | Sjælland | 23177 (15.5%) | 3398 (8.8%) | 26575 (14.1%) |
| | | Syddanmark | 24706 (16.5%) | 3139 (8.1%) | 27845 (14.8%) |
| | | Vaccination status at the time of positive test | | | |
| | | Unvaccinated | 60964 (40.7%) | 5308 (13.7%) | 66272 (35.2%) |
| | | ≥0d after first dose | 8209 (5.5%) | 1727 (4.5%) | 9936 (5.3%) |
| | | ≥14d after second dose | 49034 (32.8%) | 20407 (52.7%) | 69441 (36.9%) |
| | | ≥153d after second dose | 28151 (18.8%) | 9576 (24.7%) | 37727 (20.0%) |
| | | ≥14d after third dose | 3331 (2.2%) | 1698 (4.4%) | 5029 (2.7%) |
| | Highly desired adjustment variable | Reinfection status | | | |
| | | First infection | 147828 (98.8%) | 36570 (94.5%) | 184398 (97.9%) |
| | | Reinfection | 1861 (1.2%) | 2146 (5.5%) | 4007 (2.1%) |
| | Desired adjustment variable | Ethnicity/Country of birth | | | |
| | | Danish | 122052 (81.5%) | 33176 (85.7%) | 155228 (82.4%) |
| | | Second-generation immigrant | 8609 (5.8%) | 1299 (3.4%) | 9908 (5.3%) |
| | | First-generation immigrant | 19028 (12.7%) | 4241 (11.0%) | 23269 (12.4%) |
| | | Comorbidity | | | |
| | | No comorbidities | 125556 (83.9%) | 33765 (87.2%) | 159321 (84.6%) |
| | | ≥1 comorbidity | 24133 (16.1%) | 4951 (12.8%) | 29084 (15.4%) |
| England | | | (N=429231) | (N=914951) | (N=1344182) |
| | Required adjustment variable | Sex | | | |
| | | Female | 223332 (52.0%) | 490812 (53.6%) | 714144 (53.1%) |
| | | Male | 205899 (48.0%) | 424139 (46.4%) | 630038 (46.9%) |
| | | Age (years) | | | |
| | | Mean (SD) | 30.5 (18.5) | 35.3 (15.9) | 33.7 (16.9) |
| | | Median (Q1-Q3) | 31.0 (12.0-45.0) | 33.0 (24.0-46.0) | 33.0 (21.0-46.0) |
| | | 0-19 | 155186 (36.2%) | 133918 (14.6%) | 289104 (21.5%) |
| | | 20-39 | 122957 (28.6%) | 448159 (49.0%) | 571116 (42.5%) |
| | | 40-49 | 79133 (18.4%) | 147319 (16.1%) | 226452 (16.8%) |
| | | 50-59 | 47845 (11.1%) | 113381 (12.4%) | 161226 (12.0%) |
| | | 60-69 | 17597 (4.1%) | 52007 (5.7%) | 69604 (5.2%) |
| | | ≥70 | 6513 (1.5%) | 20167 (2.2%) | 26680 (2.0%) |
| | | Calendar week | | | |
| | | ISO week 48, 2021 | 157378 (36.7%) | 3459 (0.4%) | 160837 (12.0%) |
| | | ISO week 49, 2021 | 139431 (32.5%) | 34250 (3.7%) | 173681 (12.9%) |
| | | ISO week 50, 2021 | 81215 (18.9%) | 198634 (21.7%) | 279849 (20.8%) |
| | | ISO week 51, 2021 | 38749 (9.0%) | 361719 (39.5%) | 400468 (29.8%) |
| | | ISO week 52, 2021 | 12458 (2.9%) | 316889 (34.6%) | 329347 (24.5%) |
| | | Area of residence | | | |
| | | East of England | 48108 (11.2%) | 81063 (8.9%) | 129171 (9.6%) |
| | | London | 47528 (11.1%) | 186165 (20.3%) | 233693 (17.4%) |
| | | Midlands | 77648 (18.1%) | 143163 (15.6%) | 220811 (16.4%) |
| | | North East and Yorkshire | 79609 (18.5%) | 160866 (17.6%) | 240475 (17.9%) |
| | | North West | 58973 (13.7%) | 170838 (18.7%) | 229811 (17.1%) |
| | | South East | 72046 (16.8%) | 119925 (13.1%) | 191971 (14.3%) |
| | | South West | 45319 (10.6%) | 52931 (5.8%) | 98250 (7.3%) |
| | | Vaccination status at the time of positive test | | | |
| | | Unvaccinated | 179523 (41.8%) | 161170 (17.6%) | 340693 (25.3%) |
| | | ≥0d after first dose | 34028 (7.9%) | 60682 (6.6%) | 94710 (7.0%) |
| | | ≥14d after second dose | 60614 (14.1%) | 230801 (25.2%) | 291415 (21.7%) |
| | | ≥153d after second dose | 137944 (32.1%) | 281181 (30.7%) | 419125 (31.2%) |
| | | ≥14d after third dose | 17122 (4.0%) | 181117 (19.8%) | 198239 (14.7%) |
| | Highly desired adjustment variable | Reinfection status | | | |
| | | First infection | 423850 (98.7%) | 830188 (90.7%) | 1254038 (93.3%) |
| | | Reinfection | 5381 (1.3%) | 84763 (9.3%) | 90144 (6.7%) |
| | Desired adjustment variable | Ethnicity/Country of birth | | | |
| | | White | 363157 (84.6%) | 733624 (80.2%) | 1096781 (81.6%) |
| | | Asian | 26279 (6.1%) | 66974 (7.3%) | 93253 (6.9%) |
| | | Black | 8592 (2.0%) | 42911 (4.7%) | 51503 (3.8%) |
| | | Mixed/Other | 15217 (3.5%) | 35850 (3.9%) | 51067 (3.8%) |
| | | Unknown | 15986 (3.7%) | 35592 (3.9%) | 51578 (3.8%) |
| | | Socioeconomic status/deprivation indicator | | | |
| | | IMD decile 1 | 41076 (9.6%) | 82873 (9.1%) | 123949 (9.2%) |
| | | IMD decile 2 | 38467 (9.0%) | 92257 (10.1%) | 130724 (9.7%) |
| | | IMD decile 3 | 39691 (9.2%) | 98100 (10.7%) | 137791 (10.3%) |
| | | IMD decile 4 | 41442 (9.7%) | 93182 (10.2%) | 134624 (10.0%) |
| | | IMD decile 5 | 41687 (9.7%) | 91302 (10.0%) | 132989 (9.9%) |
| | | IMD decile 6 | 43922 (10.2%) | 91130 (10.0%) | 135052 (10.0%) |
| | | IMD decile 7 | 43863 (10.2%) | 90912 (9.9%) | 134775 (10.0%) |
| | | IMD decile 8 | 45208 (10.5%) | 93050 (10.2%) | 138258 (10.3%) |
| | | IMD decile 9 | 46155 (10.8%) | 92217 (10.1%) | 138372 (10.3%) |
| | | IMD decile 10 | 47720 (11.1%) | 89928 (9.8%) | 137648 (10.2%) |
| Luxembourg | | | (N=1286) | (N=3167) | (N=4453) |
| | Required adjustment variable | Sex | | | |
| | | Female | 639 (49.7%) | 1634 (51.6%) | 2273 (51.0%) |
| | | Male | 647 (50.3%) | 1533 (48.4%) | 2180 (49.0%) |
| | | Age (years) | | | |
| | | Mean (SD) | 36.1 (21.1) | 32.9 (20.5) | 33.8 (20.7) |
| | | Median (Q1-Q3) | 37.0 (19.0-51.0) | 32.0 (16.0-45.0) | 33.0 (17.0-47.0) |
| | | 0-19 | 326 (25.4%) | 918 (29.0%) | 1244 (27.9%) |
| | | 20-39 | 389 (30.2%) | 1118 (35.3%) | 1507 (33.8%) |
| | | 40-49 | 222 (17.3%) | 536 (16.9%) | 758 (17.0%) |
| | | 50-59 | 189 (14.7%) | 307 (9.7%) | 496 (11.1%) |
| | | 60-69 | 81 (6.3%) | 115 (3.6%) | 196 (4.4%) |
| | | ≥70 | 79 (6.1%) | 173 (5.5%) | 252 (5.7%) |
| | | Calendar week | | | |
| | | ISO week 49, 2021 | 404 (31.4%) | 5 (0.2%) | 409 (9.2%) |
| | | ISO week 50, 2021 | 482 (37.5%) | 66 (2.1%) | 548 (12.3%) |
| | | ISO week 51, 2021 | 240 (18.7%) | 255 (8.1%) | 495 (11.1%) |
| | | ISO week 52, 2021 | 87 (6.8%) | 503 (15.9%) | 590 (13.2%) |
| | | ISO week 1, 2022 | 50 (3.9%) | 682 (21.5%) | 732 (16.4%) |
| | | ISO week 2, 2022 | 17 (1.3%) | 668 (21.1%) | 685 (15.4%) |
| | | ISO week 3, 2022 | 6 (0.5%) | 988 (31.2%) | 994 (22.3%) |
| | | Area of residence | | | |
| | | Luxembourg | 1286 (100%) | 3167 (100%) | 4453 (100%) |
| | | Vaccination status at the time of positive test | | | |
| | | Unvaccinated | 683 (53.1%) | 1072 (33.8%) | 1755 (39.4%) |
| | | ≥0d after first dose | 83 (6.5%) | 271 (8.6%) | 354 (7.9%) |
| | | ≥14d after second dose | 146 (11.4%) | 458 (14.5%) | 604 (13.6%) |
| | | ≥153d after second dose | 352 (27.4%) | 948 (29.9%) | 1300 (29.2%) |
| | | ≥14d after third dose | 22 (1.7%) | 418 (13.2%) | 440 (9.9%) |
| | Highly desired adjustment variable | Reinfection status | | | |
| | | First infection | 1253 (97.4%) | 2867 (90.5%) | 4120 (92.5%) |
| | | Reinfection | 33 (2.6%) | 300 (9.5%) | 333 (7.5%) |
| | Desired adjustment variable | International travel within 14 days before positive test | | | |
| | | Yes | 50 (3.9%) | 175 (5.5%) | 225 (5.1%) |
| | | No | 1236 (96.1%) | 2992 (94.5%) | 4228 (94.9%) |
| Norway | | | (N=52091) | (N=40827) | (N=92918) |
| | Required adjustment variable | Sex | | | |
| | | Female | 25865 (49.7%) | 20297 (49.7%) | 46162 (49.7%) |
| | | Male | 26226 (50.3%) | 20530 (50.3%) | 46756 (50.3%) |
| | | Age (years) | | | |
| | | Mean (SD) | 29.9 (19.5) | 31.4 (17.0) | 30.5 (18.5) |
| | | Median (Q1-Q3) | 30.0 (11.0-44.0) | 29.0 (19.0-43.0) | 29.0 (14.0-44.0) |
| | | 0-19 | 20413 (39.2%) | 10413 (25.5%) | 30826 (33.2%) |
| | | 20-39 | 14221 (27.3%) | 17914 (43.9%) | 32135 (34.6%) |
| | | 40-49 | 8664 (16.6%) | 5811 (14.2%) | 14475 (15.6%) |
| | | 50-59 | 4864 (9.3%) | 4303 (10.5%) | 9167 (9.9%) |
| | | 60-69 | 2597 (5.0%) | 1566 (3.8%) | 4163 (4.5%) |
| | | ≥70 | 1332 (2.6%) | 820 (2.0%) | 2152 (2.3%) |
| | | Calendar week | | | |
| | | ISO week 49, 2021 | 19964 (38.3%) | 1276 (3.1%) | 21240 (22.9%) |
| | | ISO week 50, 2021 | 17016 (32.7%) | 3262 (8.0%) | 20278 (21.8%) |
| | | ISO week 51, 2021 | 8033 (15.4%) | 5045 (12.4%) | 13078 (14.1%) |
| | | ISO week 52, 2021 | 4731 (9.1%) | 11612 (28.4%) | 16343 (17.6%) |
| | | ISO week 1, 2022 | 2347 (4.5%) | 19632 (48.1%) | 21979 (23.7%) |
| | | Area of residence | | | |
| | | East Norway | 11878 (22.8%) | 6483 (15.9%) | 18361 (19.8%) |
| | | Oslo-wide area | 22998 (44.1%) | 23816 (58.3%) | 46814 (50.4%) |
| | | South and West Norway | 12906 (24.8%) | 7659 (18.8%) | 20565 (22.1%) |
| | | Mid-Norway | 3583 (6.9%) | 2262 (5.5%) | 5845 (6.3%) |
| | | North Norway | 726 (1.4%) | 607 (1.5%) | 1333 (1.4%) |
| | | Vaccination status at the time of positive test | | | |
| | | Unvaccinated | 23188 (44.5%) | 8421 (20.6%) | 31609 (34.0%) |
| | | ≥0d after first dose | 4602 (8.8%) | 3956 (9.7%) | 8558 (9.2%) |
| | | ≥14d after second dose | 16346 (31.4%) | 20442 (50.1%) | 36788 (39.6%) |
| | | ≥153d after second dose | 7124 (13.7%) | 5294 (13.0%) | 12418 (13.4%) |
| | | ≥14d after third dose | 831 (1.6%) | 2714 (6.6%) | 3545 (3.8%) |
| | Highly desired adjustment variable | Reinfection status | | | |
| | | First infection | 51889 (99.6%) | 39127 (95.8%) | 91016 (98.0%) |
| | | Reinfection | 202 (0.4%) | 1700 (4.2%) | 1902 (2.0%) |
| | Desired adjustment variable | Ethnicity/Country of birth | | | |
| | | Born in Norway | 38992 (74.9%) | 28808 (70.6%) | 67800 (73.0%) |
| | | Not born in Norway | 12367 (23.7%) | 11654 (28.5%) | 24021 (25.9%) |
| | | Unknown | 732 (1.4%) | 365 (0.9%) | 1097 (1.2%) |
| | | Socioeconomic status/deprivation indicator | | | |
| | | Not overcrowded home | 39147 (75.2%) | 30113 (73.8%) | 69260 (74.5%) |
| | | Overcrowded home | 9186 (17.6%) | 7453 (18.3%) | 16639 (17.9%) |
| | | Overcrowding status unknown | 3758 (7.2%) | 3261 (8.0%) | 7019 (7.6%) |
| | | Comorbidity | | | |
| | | No underlying comorbidities | 45994 (88.3%) | 36744 (90.0%) | 82738 (89.0%) |
| | | Medium-risk comorbidity | 5547 (10.6%) | 3740 (9.2%) | 9287 (10.0%) |
| | | High-risk comorbidity | 550 (1.1%) | 343 (0.8%) | 893 (1.0%) |
| Portugal | | | (N=9150) | (N=6469) | (N=15619) |
| | Required adjustment variable | Sex | | | |
| | | Female | 4505 (49.2%) | 3310 (51.2%) | 7815 (50.0%) |
| | | Male | 4645 (50.8%) | 3159 (48.8%) | 7804 (50.0%) |
| | | Age (years) | | | |
| | | Mean (SD) | 43.3 (15.8) | 37.1 (14.8) | 40.7 (15.7) |
| | | Median (Q1-Q3) | 42.1 (30.7-54.0) | 34.4 (24.4-47.4) | 39.6 (27.5-51.2) |
| | | 0-19 | 403 (4.4%) | 630 (9.7%) | 1033 (6.6%) |
| | | 20-39 | 3692 (40.3%) | 3250 (50.2%) | 6942 (44.4%) |
| | | 40-49 | 2100 (23.0%) | 1295 (20.0%) | 3395 (21.7%) |
| | | 50-59 | 1492 (16.3%) | 801 (12.4%) | 2293 (14.7%) |
| | | 60-69 | 945 (10.3%) | 336 (5.2%) | 1281 (8.2%) |
| | | ≥70 | 518 (5.7%) | 157 (2.4%) | 675 (4.3%) |
| | | Calendar week | | | |
| | | ISO week 48, 2021 | 296 (3.2%) | 13 (0.2%) | 309 (2.0%) |
| | | ISO week 49, 2021 | 3586 (39.2%) | 209 (3.2%) | 3795 (24.3%) |
| | | ISO week 50, 2021 | 3104 (33.9%) | 1129 (17.5%) | 4233 (27.1%) |
| | | ISO week 51, 2021 | 1942 (21.2%) | 4142 (64.0%) | 6084 (39.0%) |
| | | ISO week 52, 2021 | 222 (2.4%) | 976 (15.1%) | 1198 (7.7%) |
| | | Area of residence | | | |
| | | Algarve or Alentejo | 567 (6.2%) | 243 (3.8%) | 810 (5.2%) |
| | | Centro | 1011 (11.0%) | 372 (5.8%) | 1383 (8.9%) |
| | | Lisboa e Vale do Tejo | 1778 (19.4%) | 2298 (35.5%) | 4076 (26.1%) |
| | | Norte | 5794 (63.3%) | 3556 (55.0%) | 9350 (59.9%) |
| | | Vaccination status at the time of positive test | | | |
| | | Unvaccinated | 972 (10.6%) | 751 (11.6%) | 1723 (11.0%) |
| | | ≥0d after first dose | 1821 (19.9%) | 1208 (18.7%) | 3029 (19.4%) |
| | | ≥14d after second dose | 3288 (35.9%) | 2533 (39.2%) | 5821 (37.3%) |
| | | ≥153d after second dose | 2855 (31.2%) | 1683 (26.0%) | 4538 (29.1%) |
| | | ≥14d after third dose | 214 (2.3%) | 294 (4.5%) | 508 (3.3%) |
| | Highly desired adjustment variable | Reinfection status | | | |
| | | First infection | 9011 (98.5%) | 6023 (93.1%) | 15034 (96.3%) |
| | | Reinfection | 139 (1.5%) | 446 (6.9%) | 585 (3.7%) |
| | Desired adjustment variable | Ethnicity/Country of birth | | | |
| | | Portugal | 8879 (97.0%) | 6289 (97.2%) | 15168 (97.1%) |
| | | Other | 271 (3.0%) | 180 (2.8%) | 451 (2.9%) |
| Scotland | | | (N=35534) | (N=47744) | (N=83278) |
| | Required adjustment variable | Sex | | | |
| | | Male | 17217 (48.5%) | 21442 (44.9%) | 38659 (46.4%) |
| | | Female | 18317 (51.5%) | 26302 (55.1%) | 44619 (53.6%) |
| | | Age (years) | | | |
| | | Mean (SD) | 33.3 (19.5) | 33.2 (17.0) | 33.3 (18.1) |
| | | Median (Q1-Q3) | 35.0 (13.0-49.0) | 32.0 (21.0-45.0) | 33.0 (18.0-47.0) |
| | | 0-19 | 11530 (32.4%) | 10860 (22.7%) | 22390 (26.9%) |
| | | 20-39 | 9116 (25.7%) | 20211 (42.3%) | 29327 (35.2%) |
| | | 40-49 | 6155 (17.3%) | 7292 (15.3%) | 13447 (16.1%) |
| | | 50-59 | 5311 (14.9%) | 5731 (12.0%) | 11042 (13.3%) |
| | | 60-69 | 2743 (7.7%) | 2918 (6.1%) | 5661 (6.8%) |
| | | ≥70 | 679 (1.9%) | 732 (1.5%) | 1411 (1.7%) |
| | | Calendar week | | | |
| | | ISO week 41, 2021 | 1778 (5.0%) | 10 (0.0%) | 1788 (2.1%) |
| | | ISO week 42, 2021 | 2495 (7.0%) | 16 (0.0%) | 2511 (3.0%) |
| | | ISO week 43, 2021 | 4407 (12.4%) | 12 (0.0%) | 4419 (5.3%) |
| | | ISO week 44, 2021 | 4031 (11.3%) | 15 (0.0%) | 4046 (4.9%) |
| | | ISO week 45, 2021 | 3578 (10.1%) | 39 (0.1%) | 3617 (4.3%) |
| | | ISO week 46, 2021 | 3475 (9.8%) | 42 (0.1%) | 3517 (4.2%) |
| | | ISO week 47, 2021 | 5514 (15.5%) | 80 (0.2%) | 5594 (6.7%) |
| | | ISO week 48, 2021 | 3500 (9.8%) | 234 (0.5%) | 3734 (4.5%) |
| | | ISO week 49, 2021 | 4163 (11.7%) | 1221 (2.6%) | 5384 (6.5%) |
| | | ISO week 50, 2021 | 1662 (4.7%) | 2484 (5.2%) | 4146 (5.0%) |
| | | ISO week 51, 2021 | 471 (1.3%) | 2447 (5.1%) | 2918 (3.5%) |
| | | ISO week 52, 2021 | 302 (0.8%) | 5927 (12.4%) | 6229 (7.5%) |
| | | ISO week 1, 2022 | 90 (0.3%) | 4352 (9.1%) | 4442 (5.3%) |
| | | ISO week 2, 2022 | 37 (0.1%) | 4025 (8.4%) | 4062 (4.9%) |
| | | ISO week 3, 2022 | 16 (0.0%) | 5533 (11.6%) | 5549 (6.7%) |
| | | ISO week 4, 2022 | 4 (0.0%) | 3361 (7.0%) | 3365 (4.0%) |
| | | ISO week 5, 2022 | 9 (0.0%) | 6683 (14.0%) | 6692 (8.0%) |
| | | ISO week 6, 2022 | 2 (0.0%) | 11263 (23.6%) | 11265 (13.5%) |
| | | Area of residence | | | |
| | | Scotland | 35534 (100%) | 47744 (100%) | 83278 (100%) |
| | | Vaccination status at the time of positive test | | | |
| | | Unvaccinated | 12855 (36.2%) | 11699 (24.5%) | 24554 (29.5%) |
| | | ≥0d after first dose | 2961 (8.3%) | 3180 (6.7%) | 6141 (7.4%) |
| | | ≥14d after second dose | 9154 (25.8%) | 6788 (14.2%) | 15942 (19.1%) |
| | | ≥153d after second dose | 9723 (27.4%) | 8508 (17.8%) | 18231 (21.9%) |
| | | ≥14d after third dose | 841 (2.4%) | 17569 (36.8%) | 18410 (22.1%) |
| | Highly desired adjustment variable | Reinfection status | | | |
| | | First infection | 35278 (99.3%) | 43713 (91.6%) | 78991 (94.9%) |

| | | | | |
|---|---|---|---|---|
| | Reinfection | 256 (0.7%) | 4031 (8.4%) | 4287 (5.1%) |
| Desired adjustment variable | Ethnicity/Country of birth | | | |
| | White | 32775 (92.2%) | 43215 (90.5%) | 75990 (91.2%) |
| | Asian | 744 (2.1%) | 1635 (3.4%) | 2379 (2.9%) |
| | Black | 199 (0.6%) | 438 (0.9%) | 637 (0.8%) |
| | Mixed/Multiple ethnic group | 348 (1.0%) | 473 (1.0%) | 821 (1.0%) |
| | Other ethnic group | 108 (0.3%) | 233 (0.5%) | 341 (0.4%) |
| | Unknown | 1360 (3.8%) | 1750 (3.7%) | 3110 (3.7%) |
| | Socioeconomic status/deprivation indicator | | | |
| | IMD quintile 1 | 6773 (19.1%) | 10524 (22.0%) | 17297 (20.8%) |
| | IMD quintile 2 | 6969 (19.6%) | 9995 (20.9%) | 16964 (20.4%) |
| | IMD quintile 3 | 6763 (19.0%) | 8661 (18.1%) | 15424 (18.5%) |
| | IMD quintile 4 | 7486 (21.1%) | 9337 (19.6%) | 16823 (20.2%) |
| | IMD quintile 5 | 7543 (21.2%) | 9227 (19.3%) | 16770 (20.1%) |

**Supplementary Figure S7.** Unadjusted hazard ratios for COVID-19 cases infected with the Omicron versus Delta variants, of (A) hospital admission (COVID-19-specific, where available, or otherwise due to any cause), (B) ICU admission (COVID-19-specific), or (C) death (COVID-19-specific, where available, or otherwise due to any cause).

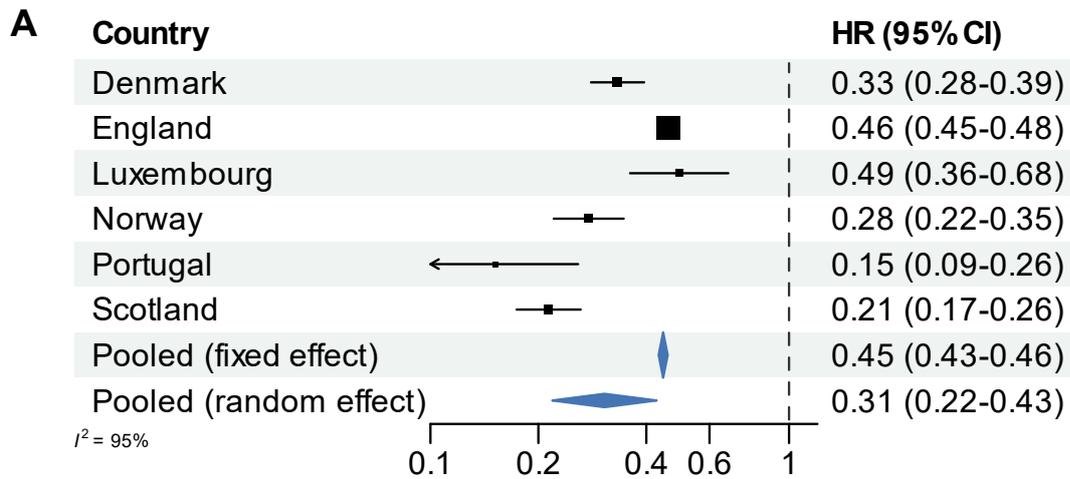

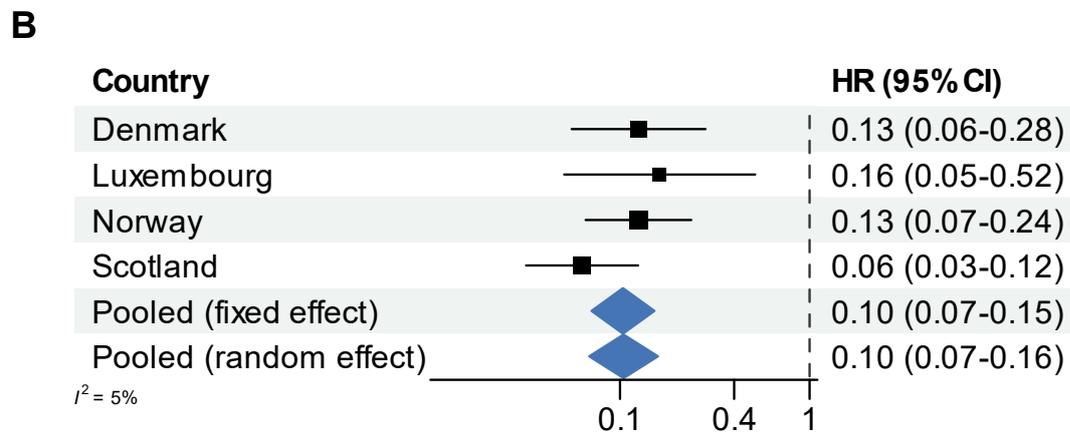

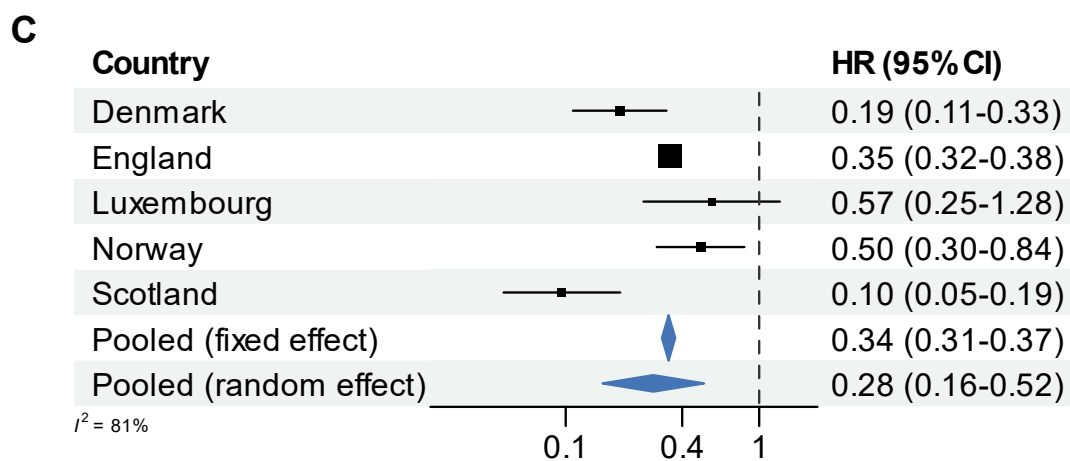

**Supplementary Figure S8:** Hazard ratios of ICU admission (COVID-19-specific) for COVID-19 cases infected with the Omicron versus Delta variants, as Figure 3F (adjusted for the required, highly desired and desired set of adjustment variables, using stratification for calendar week and regression adjustment for exact calendar date), by age group.

*Note:* Some countries observed no ICU admissions in some age groups, and were therefore not included in all age group analyses: this included Denmark for cases aged 0-19 and 40-59 years, Luxembourg for cases aged <40 years, Norway for cases aged 40-49 or 60-69 years, and Scotland for cases aged <40 years. The results for these countries and age groups were therefore not included in the sub-analyses.

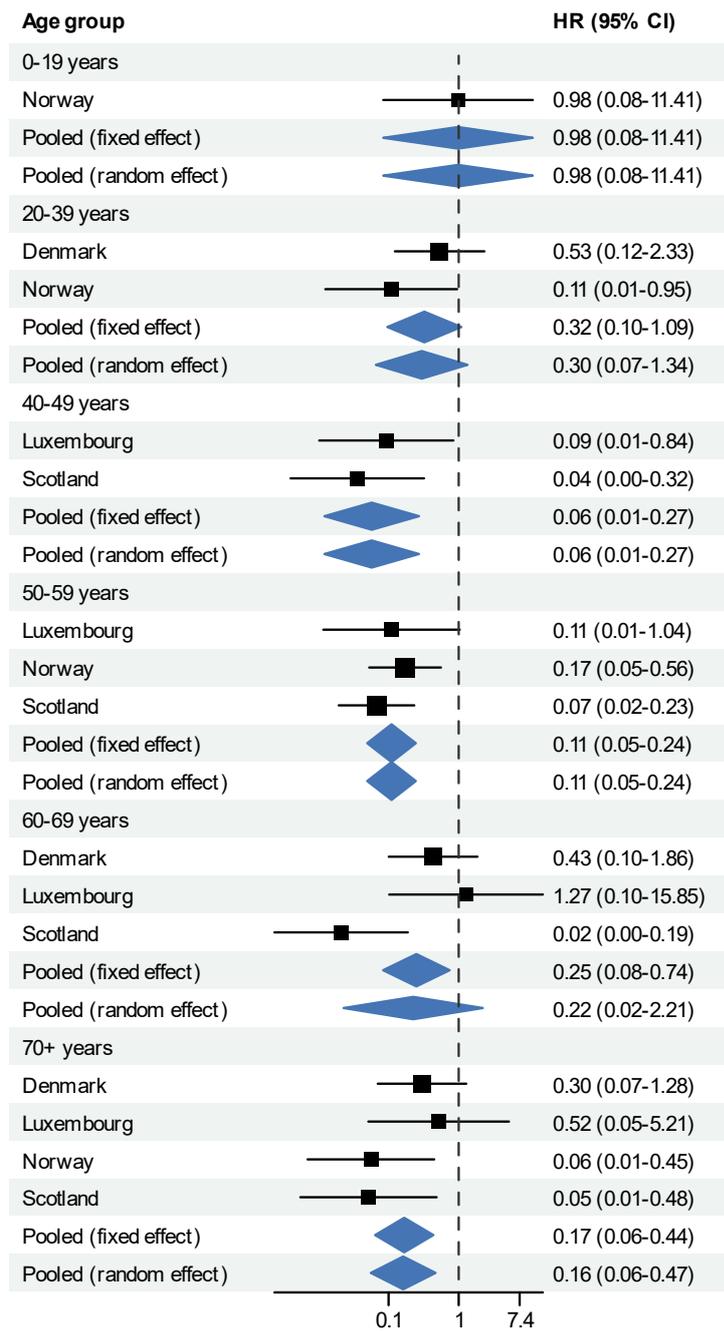

**Supplementary Figure S9:** Hazard ratios of death (COVID-19-specific, where available, or otherwise due to any cause) for COVID-19 cases infected with the Omicron versus Delta variants, as Figure 4F (adjusted for the required, highly desired and desired set of adjustment variables, using stratification for calendar week and regression adjustment for exact calendar date), by age group.

*Note:* Some countries observed no deaths in some age groups, and were therefore not included in all age group analyses: this included Denmark for cases aged <60 years, Luxembourg for cases aged <60 years, Norway for cases aged <50 years, and Scotland for cases aged <40 years. The results for these countries and age groups were therefore not included in the sub-analyses.

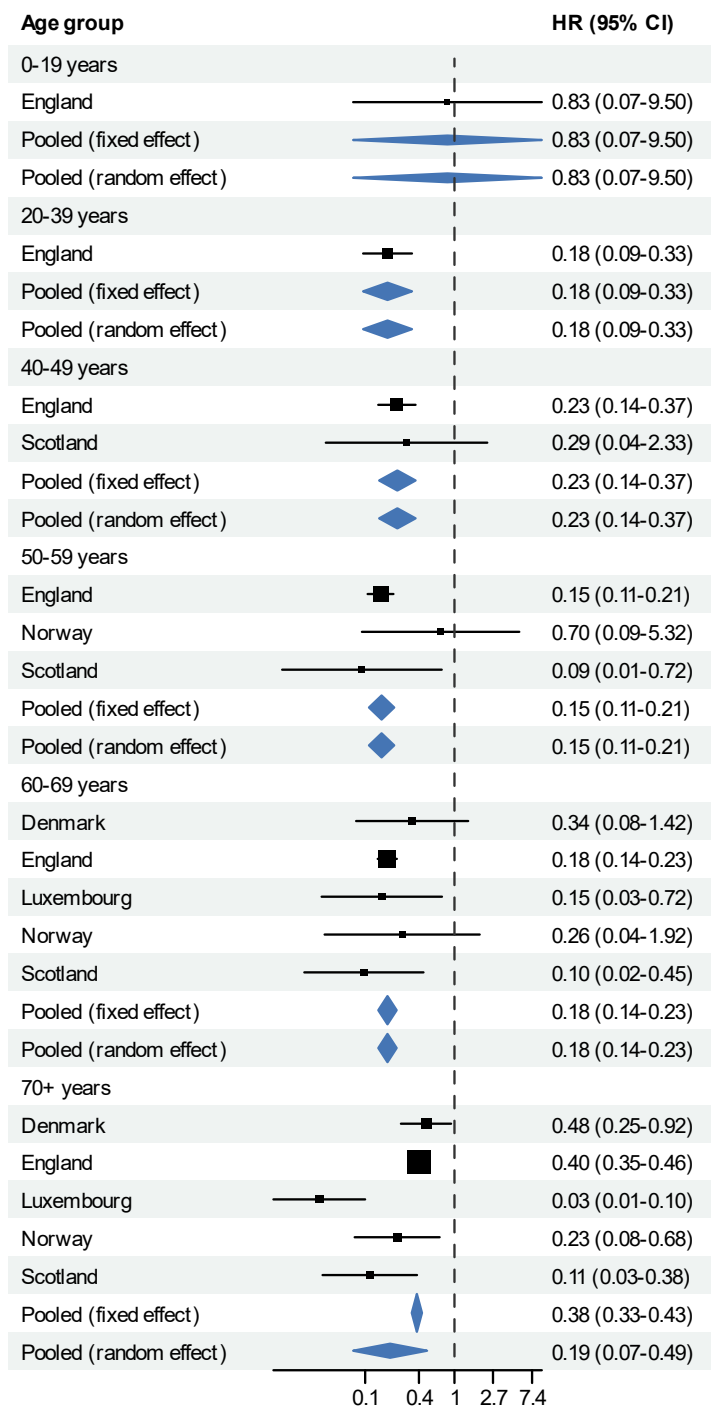

**Supplementary Figure S10:** Hazard ratios of ICU admission (COVID-19-specific) for COVID-19 cases infected with the Omicron versus Delta variants, as Figure 3F (adjusted for the required, highly desired and desired set of adjustment variables, using stratification for calendar week and regression adjustment for exact calendar date), by vaccination status.

*Note:* Some countries observed no ICU admissions in some vaccination subgroups, and were therefore not included in all vaccination status subgroup analyses.

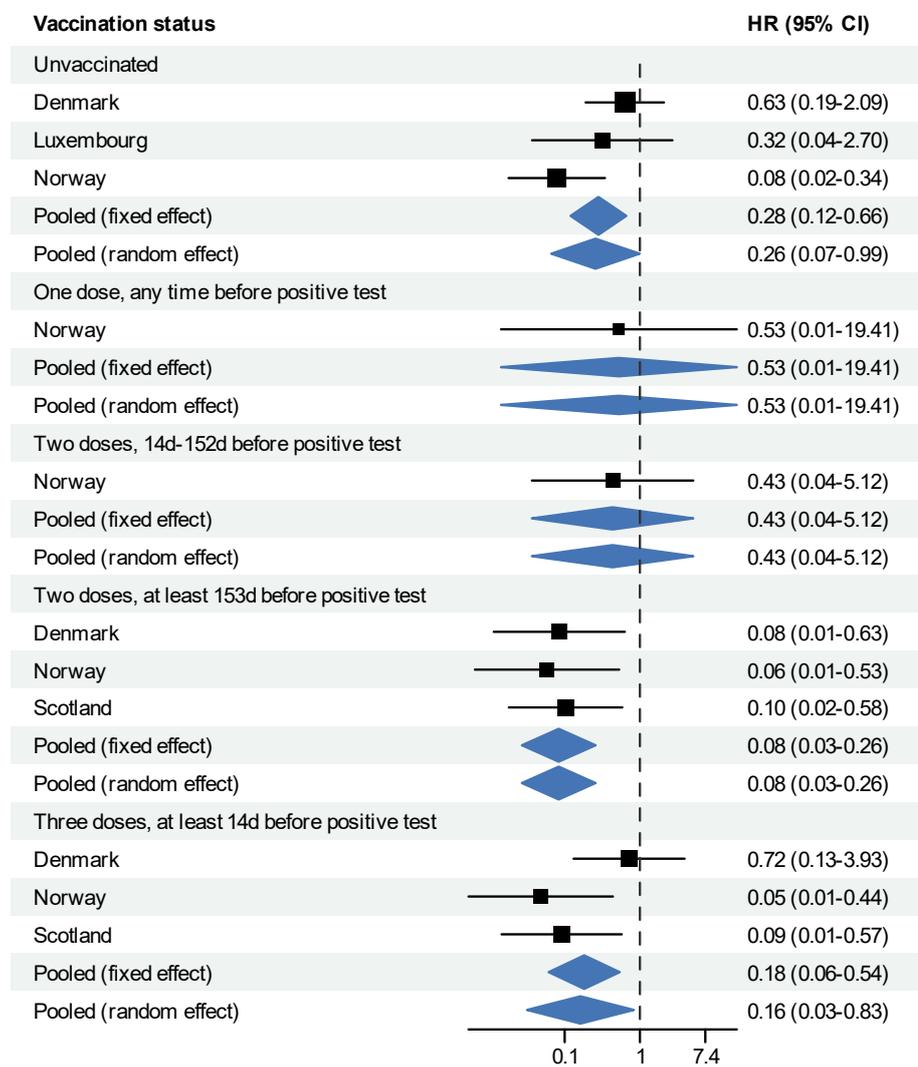

**Supplementary Figure S11:** Hazard ratios of death (COVID-19-specific, where available, or otherwise due to any cause) for COVID-19 cases infected with the Omicron versus Delta variants, as Figure 4F (adjusted for the required, highly desired and desired set of adjustment variables, using stratification for calendar week and regression adjustment for exact calendar date), by vaccination status.

*Note:* Some countries observed no deaths in some vaccination subgroups, and were therefore not included in all vaccination status subgroup analyses.

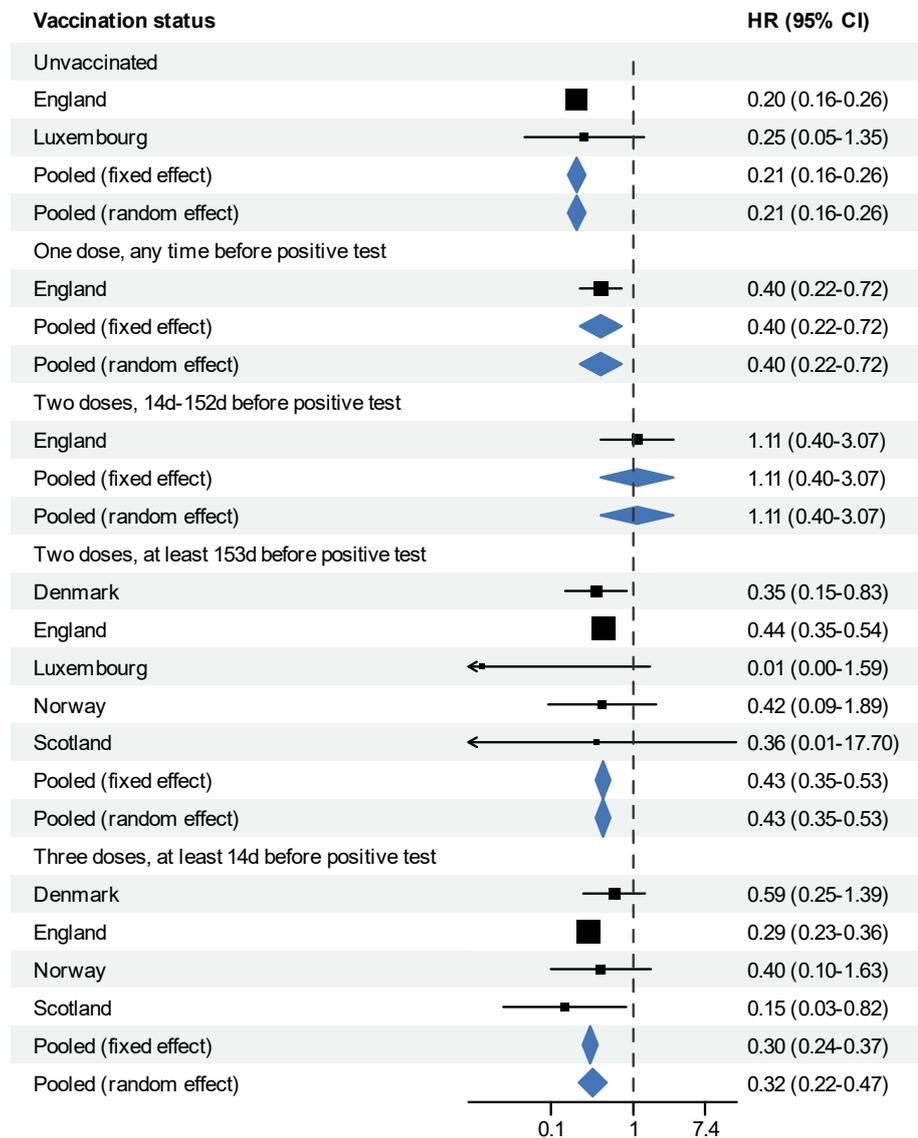

| Vaccination status | HR (95% CI) |
|---|---|
| **Unvaccinated** | |
| England | 0.20 (0.16-0.26) |
| Luxembourg | 0.25 (0.05-1.35) |
| Pooled (fixed effect) | 0.21 (0.16-0.26) |
| Pooled (random effect) | 0.21 (0.16-0.26) |
| **One dose, any time before positive test** | |
| England | 0.40 (0.22-0.72) |
| Pooled (fixed effect) | 0.40 (0.22-0.72) |
| Pooled (random effect) | 0.40 (0.22-0.72) |
| **Two doses, 14d-152d before positive test** | |
| England | 1.11 (0.40-3.07) |
| Pooled (fixed effect) | 1.11 (0.40-3.07) |
| Pooled (random effect) | 1.11 (0.40-3.07) |
| **Two doses, at least 153d before positive test** | |
| Denmark | 0.35 (0.15-0.83) |
| England | 0.44 (0.35-0.54) |
| Luxembourg | 0.01 (0.00-1.59) |
| Norway | 0.42 (0.09-1.89) |
| Scotland | 0.36 (0.01-17.70) |
| Pooled (fixed effect) | 0.43 (0.35-0.53) |
| Pooled (random effect) | 0.43 (0.35-0.53) |
| **Three doses, at least 14d before positive test** | |
| Denmark | 0.59 (0.25-1.39) |
| England | 0.29 (0.23-0.36) |
| Norway | 0.40 (0.10-1.63) |
| Scotland | 0.15 (0.03-0.82) |
| Pooled (fixed effect) | 0.30 (0.24-0.37) |
| Pooled (random effect) | 0.32 (0.22-0.47) |

**Supplementary Figure S12:** Sensitivity analysis to assess the effect of epidemic phase bias [5]. Hazard ratios of hospital admission (COVID-19-specific, where available, or otherwise due to any cause) for COVID-19 cases infected with the Omicron versus Delta variants, as Figure 2F (adjusted for the required, highly desired and desired set of adjustment variables, using stratification for calendar week and regression adjustment for exact calendar date), after adjusting for the effect of epidemic phase bias.

Epidemic phase bias may occur when comparing two virus variants that are in different phases of incidence growth or decline. When the incidence of a variant is growing, cases who test positive with the variant tend to include a higher proportion of individuals with relatively shorter time since infection, and cases who test positive with a variant whose incidence is declining tend to include individuals with relatively longer time since infection. This might bias estimates of the relative risk between the variants when date of infection is unknown and adjustments are instead based on date of positive test, if infection severity if associated with shorter times from infection to test. To assess the potential impact of such bias, Seaman and colleagues [5] proposed a sensitivity analysis where a proxy adjustment date for date of infection is constructed based on assumed range for the mean difference in time from infection to positive test between those who do not experience severe disease and those who do.

The plot shows the results assuming a range of the mean difference of between 0 days and 4 days.

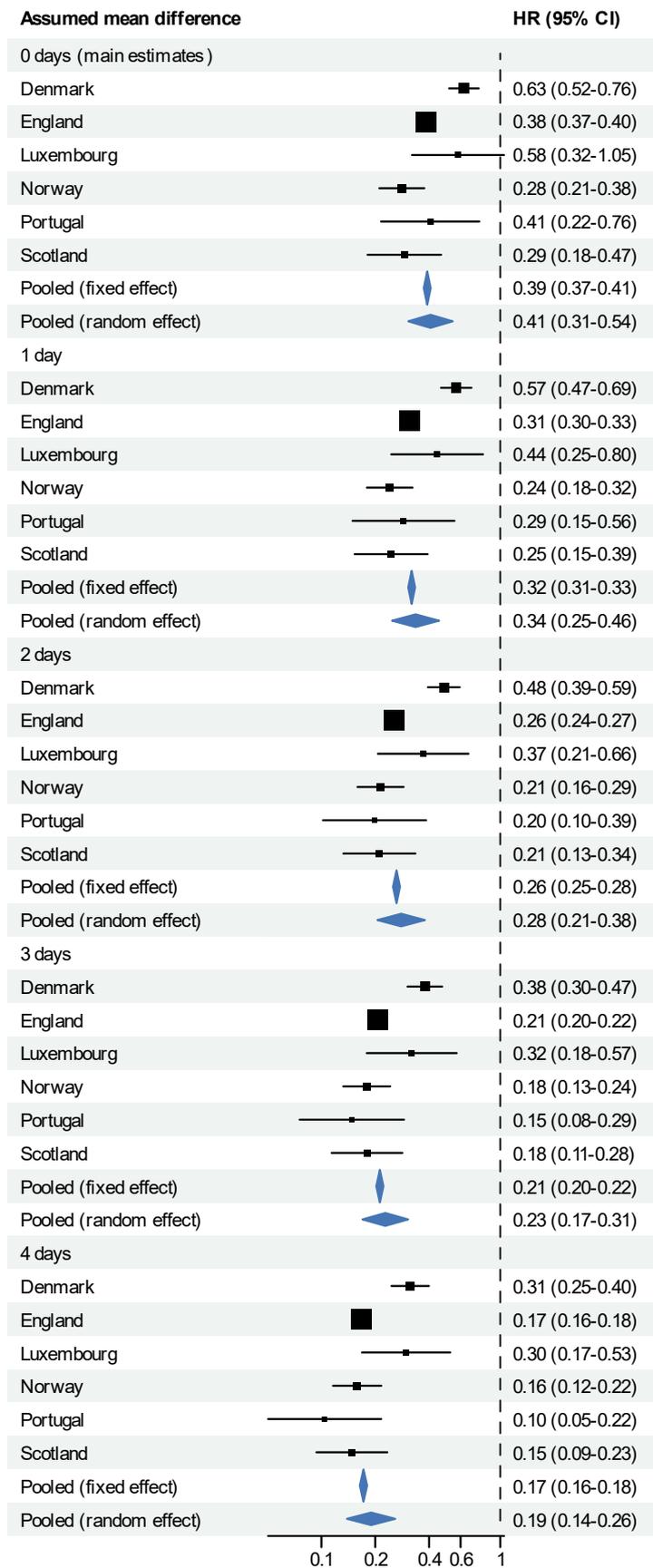

**Figure S13:** Hazard ratios of hospital admission (COVID-19-specific, where available, or otherwise due to any cause) for COVID-19 cases infected with the Omicron versus Delta variants, by country and pooled across countries using a fixed effect and a random effect model; as Figure 2, but restricted to protocol-compliant cases with reported Delta variant information in Denmark (see Methods).

(A). Adjusted for the required set of adjustment variables, using stratification for exact calendar date.
(B). Adjusted for the required and highly desired set of adjustment variables, using stratification for exact calendar date.
(C). Adjusted for the required, highly desired and desired set of adjustment variables, using stratification for exact calendar date.
(D). As (A), but using stratification for calendar week and regression adjustment for exact calendar date.
(E). As (B), but using stratification for calendar week and regression adjustment for exact calendar date.
(F). As (C), but using stratification for calendar week and regression adjustment for exact calendar date.

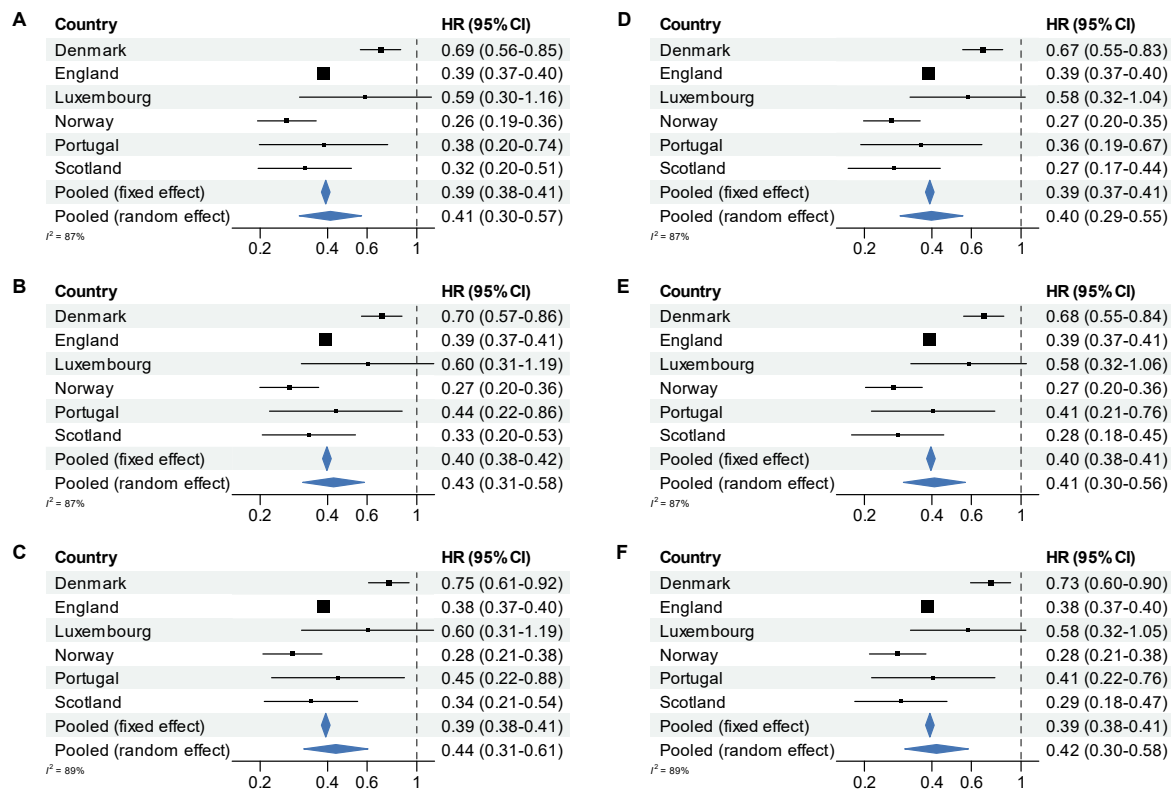

# References, Supplement B